\pdfoutput=1
\RequirePackage{lineno}
\documentclass{pj}
\usepackage{graphicx}
\usepackage{subfigure}
\usepackage{epstopdf}
\usepackage{rotating}
\usepackage[stable]{footmisc}

\newcommand{\mitlns}{Laboratory for Nuclear Science, Massachusetts Institute of Technology, Cambridge, MA}
\newcommand{\unc}{Physics Department, University of North Carolina at Chapel Hill, Chapel Hill, NC}
\newcommand{\mitmat}{Department of Materials Science and Engineering, Massachusetts Institute of Technology, Cambridge, MA}
\newcommand{\kit}{Karlsruhe Institute of Technology, Karlsruhe, Germany}
\newcommand{\boskovic}{Theoretical Physics Division, Rudjer Bo\v{s}kovi\'{c} Institute, P.O.B. 180, HR-10002 Zagreb, Croatia}
\newcommand{\artes}{Artes Calculi,Fra Grge Marti\'{c}a 24, Zagreb, Croatia}
\newcommand{\kfki}{KFKI, RMKI, Budapest, Hungary}

\begin{document}
\setcounter{page}{1}
\pjheader{ }

\title{Solving for Micro- and Macro- Scale Electrostatic Configurations Using the Robin Hood Algorithm~}
\author{\bf J.~A.~Formaggio$^{1}$\footnote{Corresponding Author: J. A. Formaggio (josephf@mit.edu).}, P. Lazi\'{c}$^{2}$, T. J. Corona$^{3}$, H. \v{S}tefan\v{c}i\'{c}$^{4}$\footnote{The work of Hrvoje \v{S}tefan\v{c}i\'{c} on this paper was performed outside his working hours at Rudjer Bo\v{s}kovi\'{c} Institute and outside facilities of Rudjer Bo\v{s}kovi\'{c} Institute.}, H.~Abraham$^{5}$, and F.~Gl\"{u}ck}$^{6,7}$\\

\noindent $^{1}$\address{\mitlns}\\

\noindent $^{2}$\address{\mitmat}\\

\noindent $^{3}$\address{\unc}\\

\noindent $^{4}$\address{\boskovic}\\

\noindent $^{5}$\address{\artes}\\

\noindent $^{6}$\address{\kit}\\

\noindent $^{7}$\address{\kfki}\\

\runningauthor{Formaggio et al.}
\tocauthor{J.~Formaggio}

\begin{abstract}  
We present a novel technique by which highly-segmented electrostatic configurations can be solved.  The Robin Hood method is a matrix-inversion algorithm optimized for solving high density boundary element method (BEM) problems. We illustrate the capabilities of this solver by studying two distinct geometry scales:  (a) the electrostatic potential of a large volume beta-detector and (b) the field enhancement present at surface of electrode nano-structures.  Geometries with elements numbering in the ${\cal O}(10^5)$ are easily modeled and solved without loss of accuracy.  The technique has recently been expanded so as to include dielectrics and magnetic materials.  
\end{abstract}

\date{\today}

\section{Introduction}
\label{sec:intro}

Numerical modeling of the electromagnetic properties of complex systems is often a necessity for many scientific and engineering disciplines.  Knowledge of the electric and magnetic fields often provides valuable insight into the system's performance, optimization parameters, and inherent response.  As these systems grow in size and complexity, so do the demands upon the model employed.  Even within a well-understood framework such as Maxwell's electromagnetic theory, the user is often confronted with a choice between accurate, model-independent solutions or timely results.  

The development of computational power in the last decades has resulted in a large number of available efficient methods for the solution of these potential problems in electrostatics, as well as in other areas of physical, engineering and technological applications. According to the basic approach towards solving electrostatic problems, methods can be roughly divided into either finite element methods (FEM), boundary element methods (BEM) and finite difference (FD) methods.  BEM methods in particular have risen in wide-spread use over recent decades~\cite{BEM,Hall,Cartwright,Gaul,Brebbia,Brebbia2}, as it often provides the most accurate means to describe a given system.  Employing BEM from a practical standpoint, however, becomes somewhat difficult for highly segmented geometries, as the requirements on computational memory grows quadratically with the number of sub-elements.  

In this article, we describe how a new BEM-solving algorithm, called Robin Hood, elegantly solves the memory problem, providing accurate results with reasonable computation times.  The technique was first introduced in 2006~\cite{RH1, RH2}, and was recently expanded to include dielectric and magnetic materials.  We illustrate its applicability and scope by studying both micro- and macro- scale systems.  Section~\ref{sec:BEM} describes the nature of the problem to be solved.  Section~\ref{sec:Technique} describes the Robin Hood method and its basic performance characteristics.  Section~\ref{sec:systems} describes its use in solving real-world systems.  We conclude in Section~\ref{sec:conclusion} with a summary and outlook for future applications.

\section{The Electrostatic Boundary Value Problem}
\label{sec:BEM}

The Boundary Element Method (BEM) is a computational technique for solving linear partial differential equations.  Compared to other popular methods (such as the Finite Element and Finite Difference Methods \cite{Szilagyi}) designed to accomplish the same goal, the BEM differs in many respects, favoring its use in an important subset of problems.  Instead of discretizing the entire region of interest, the main technique of the BEM is to discretize only the surfaces of the geometries in the region.  This effectively reduces the dimensionality of the problem and facilitates the calculation of fields for regions that extend out to infinity (rather than restricting computation to a finite region)~\cite{Poljak}. These two features make the BEM faster and more versatile than competing methods when it is applicable.

\begin{figure}[htb]
\begin{center}
{
\includegraphics[width=0.5\linewidth,clip=true]{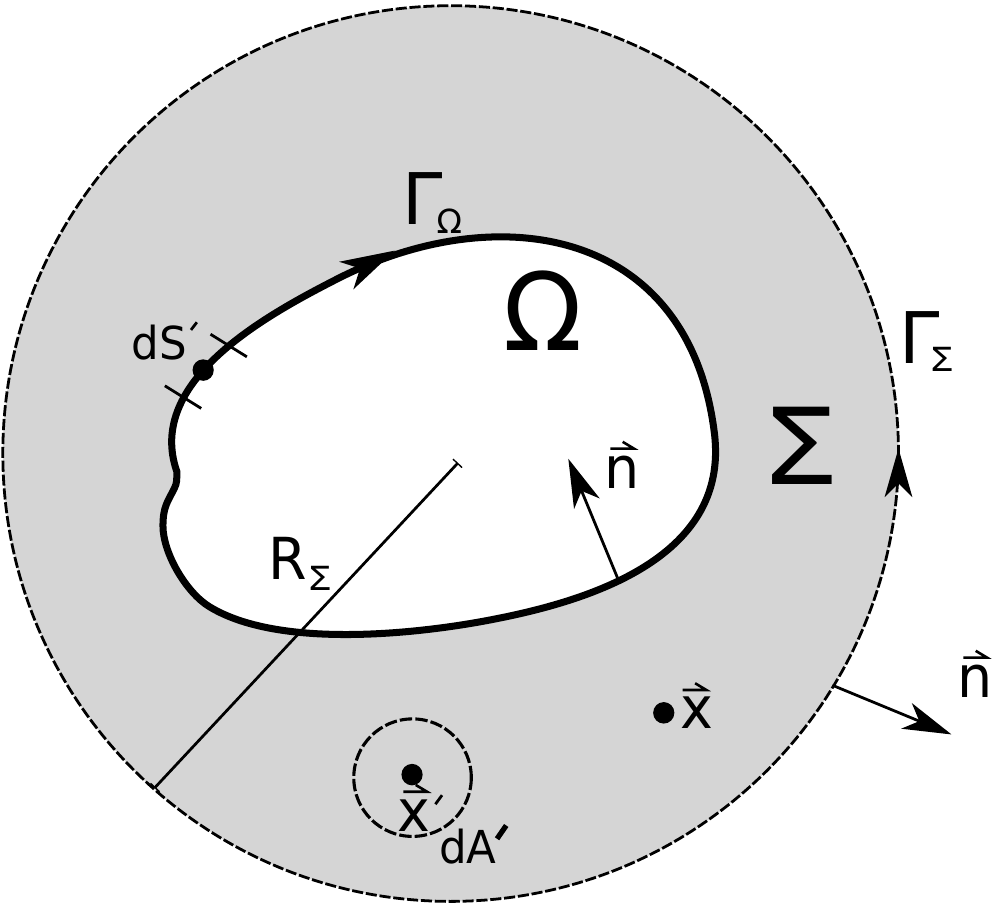}
}
\end{center}
\caption{A two-dimensional graphical depiction of the regions $\Omega$ (in white) and $\Sigma$ (in grey) (see Fig.~\ref{bemBoundingGeometry}).  The vector $\vec{n}$ describes the unit normal  to the boundaries ($\Gamma_{\Omega}$ and $\Gamma_{\Sigma}$) of these regions, and $\vec{x}$ is the observation point.}
\label{bemBoundingGeometry}
\end{figure}

We begin by defining a two-dimensional inner region $\Omega$, bounded by a piecewise smooth contour $\Gamma_\Omega$ with clockwise orientation.  The region $\Omega$ is bounded by an external region $\Sigma$, which is bounded by a circular, counter-clockwise contour $\Gamma_\Sigma$ of radius $R_\Sigma$ and internally by the contour $\Gamma_\Omega$. To derive the principle equation of the BEM, one begins by applying Green's second identity to a region $\Sigma$ encapsulated by mixed (Dirichlet and Neumann) boundaries $\Gamma_\Omega$ and $\Gamma_\Sigma$, for two arbitrary scalar functions $U$ and $W$~\cite{Jackson}: 

\begin{eqnarray}
\int_\Sigma \left( U \nabla ^2 W - W \nabla^2 U \right) dA^\prime = \qquad \qquad \qquad & & \nonumber \\
 \int_{\Gamma_\Sigma + \Gamma_\Omega} \left( U \frac{\partial W}{\partial n} - W \frac{\partial U}{\partial n} \right) dS^\prime, &&
\label{eq:Greens2ndEq}
\end{eqnarray}

We define the electric potential, $U(\vec{x})$ in Eq. \ref{eq:Greens2ndEq} such that the Laplace relation is satisfied, i.e.:

\begin{equation}
\nabla^2U(\vec{x}) = 0.
\label{eq:laplaceCondition}
\end{equation}

Defining $W$ as the fundamental solution to the Laplace equation, we find: 

\begin{equation}
W = G(\vec{x},\vec{x}^\prime) = \frac{1}{4 \pi \epsilon_0} \frac{1}{|\vec{x}-\vec{x}^\prime|}.
\label{eq:GreensFunction}
\end{equation}

\begin{figure}[htb]
\begin{center}
{
\includegraphics[width=0.5\linewidth,clip=true]{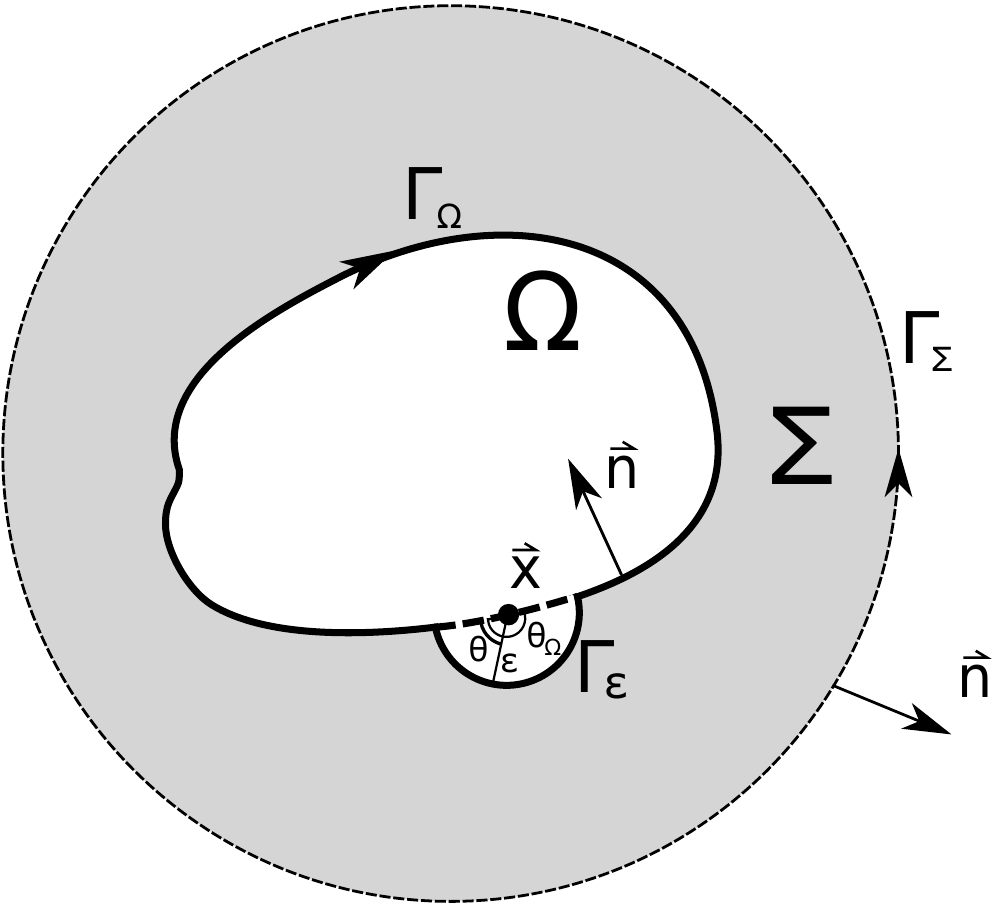}
}
\end{center}
\caption{Same as Fig.~\ref{bemBoundingGeometry} except the $\Gamma_\Omega$ boundary is deformed to include a small circular segment of radius $\epsilon$ to avoid the singularity of $\nabla^2 W(\vec{r})$ at the boundary. In the limit $\epsilon \rightarrow 0$, the original boundary is recovered.}
\label{bemBoundingLimit}
\end{figure}

When extending $\Gamma_\Sigma$ out to infinity and performing contour integrations about the boundaries of $\Sigma$, one is left with the relation:

\begin{equation}
c_1(\vec{x}) \cdot U(\vec{x}) = \int _{\Gamma_{\Omega}} \left( U \frac{\partial G}{\partial n} - G \frac{\partial U}{\partial n} \right) dS^\prime,
\label{greenwithlaplacefinal}
\end{equation}
\begin{equation}
c_1(\vec{x}) = \left\{ \begin{array}{ll}
1 & \mbox{$\vec{x} \in \Sigma$} \\
1 - \frac{\theta_{\Omega}}{2 \pi} & \mbox{$\vec{x} \in \Gamma_{\Omega}$} \\
0 & \mbox{$\vec{x} \notin \Sigma$}
\end{array} \right. ,
\label{greenwithlaplacefinalconditions}
\end{equation}

\noindent where $\theta_\Omega$ is a boundary angle of integration whose pivot is located at $\Gamma_\Omega$ (see Fig.~\ref{bemBoundingLimit}).

Often, Eq. \ref{greenwithlaplacefinal} is numerically solved by discretizing the boundary $\Gamma_\Omega$ (this technique is known as the \emph{direct BEM}).  Another technique, known as the \emph{indirect BEM}\footnote{This technique is also known as the Source Element Method (SEM), Source Integration Method (SIM), and Charge-Density (Integral) Method.}, applies the above formalism to the inner region $\Omega$ (as in Fig. \ref{bemBoundingGeometryInternal}), yielding a similar relation to Eq. \ref{greenwithlaplacefinal}: 

\begin{figure}[htb]
\begin{center}
{
\includegraphics[width=0.5\linewidth,clip=true]{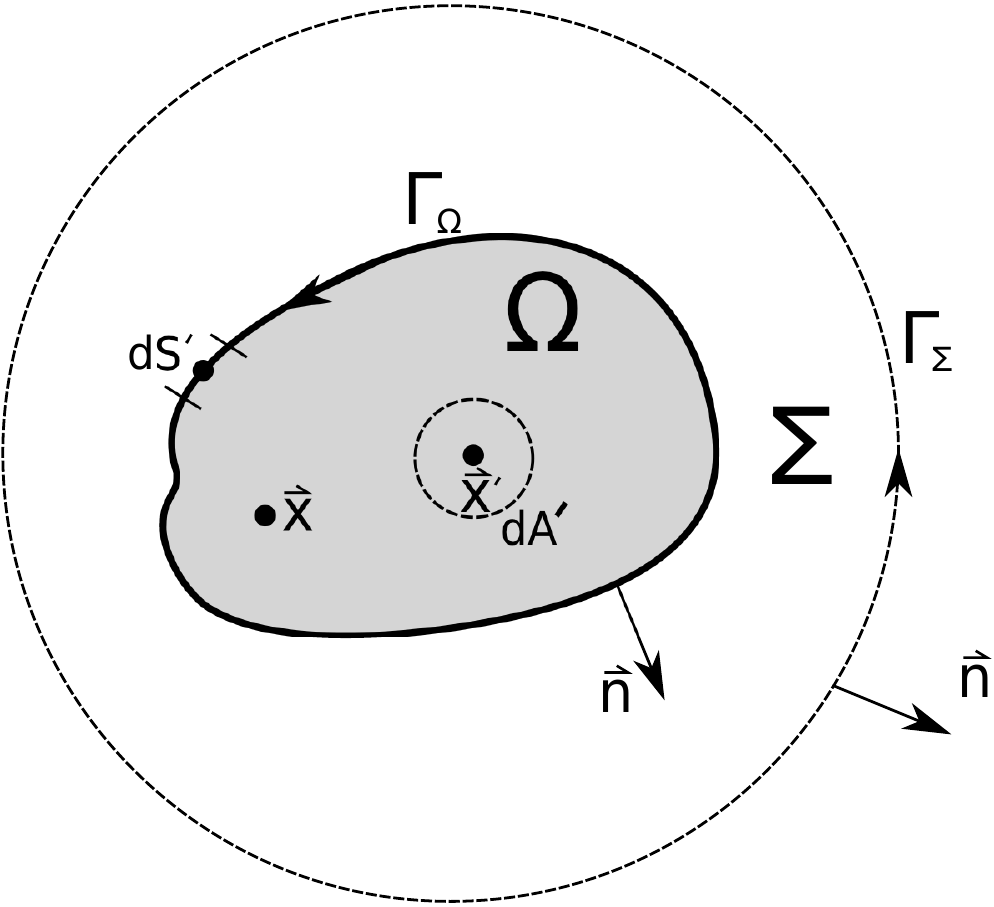}
}
\end{center}
\caption{The equivalent figure to Fig. \ref{bemBoundingGeometry} for the internal region $\Omega$.  The shaded region denotes the domain of interest for computation.}
\label{bemBoundingGeometryInternal}
\end{figure}

\begin{equation}
c_2(\vec{x}) \cdot \tilde{U}(\vec{x}) = \int _{\Gamma_{\Omega}} \left( \tilde{U} \frac{\partial G}{\partial n} - G \frac{\partial \tilde{U}}{\partial n} \right) dS^\prime,
\label{greenwithlaplaceinterior}
\end{equation}
\begin{equation}
c_2(\vec{x}) = \left\{ \begin{array}{ll}
1 & \mbox{$\vec{x} \in \Omega$} \\
\frac{\theta_{\Omega}}{2 \pi} & \mbox{$\vec{x} \in \Gamma_{\Omega}$} \\
0 & \mbox{$\vec{x} \notin \Omega$}
\end{array} \right. ,
\label{greenwithlaplaceinteriorconditions}
\end{equation}

\noindent which differs from Equation \ref{greenwithlaplacefinal} by the definition of the boundary angle, the boundary's orientation and the direction of the boundary normal.  By reversing the boundary normal in Eq. \ref{greenwithlaplaceinterior} and restricting the domain to $\Sigma$, Eq. \ref{greenwithlaplaceinterior} simplifies to

\begin{equation}
0 = \int _{\Gamma_{\Omega}} \left( \tilde{U} \frac{\partial G}{\partial n} + G \frac{\partial \tilde{U}}{\partial n} \right) dS^\prime.
\label{greenwithlaplaceinteriorrflip}
\end{equation}

By matching the boundary conditions and summing Eqs. \ref{greenwithlaplacefinal} and \ref{greenwithlaplaceinteriorrflip}, one finds

\begin{equation}
U(\vec{x}) = \int_{\Gamma_{\Omega}} G \left( \frac{\partial U}{\partial n} + \frac{\partial \tilde{U}}{\partial n} \right) dS^\prime.
\label{suminteriorexterior}
\end{equation}

Defining $\sigma$ to be the sum of the fluxes across $\Gamma_\Omega$:

\begin{equation}
\sigma = \left( \frac{\partial U}{\partial n} + \frac{\partial \tilde{U}}{\partial n} \right),
\label{sumOfFluxes}
\end{equation}

one arrives at the final form of the indirect BEM: 

\begin{equation}
U(\vec{x}) = \int_{\Gamma_\Omega} G(\vec{x},\vec{x}^\prime) \sigma(\vec{x}^\prime) dS^\prime. 
\label{indirectBEMFinalForm}
\end{equation}

\begin{figure}[htb]
\begin{center}
{
\includegraphics[width=0.32\linewidth,clip=true]{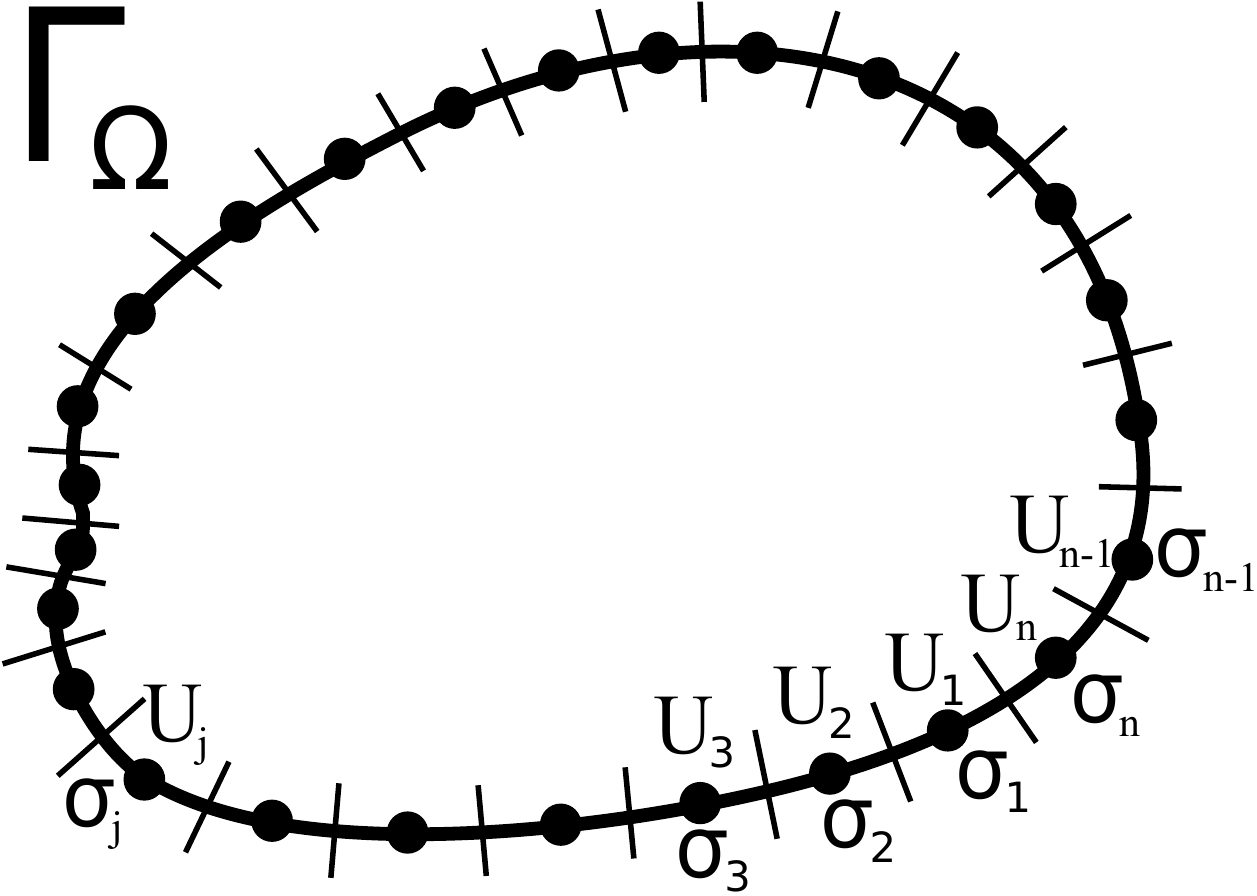}
}
\end{center}
\caption{Discretization of $\Gamma_{\Omega}$ into $n$ sub-elements for numerical computation.  Charge densities are constant along a sub-element.}
\label{bemBoundingGeometryDiscrete}
\end{figure}

If the distribution of charges is known {\it apriori}, such a calculation would be relatively trivial to solve.  Of course, the distribution of charges is {\em not} known, but needs to be solved for the given geometry.  What is typically known, however, are the boundary conditions of the system.  In the case of conducting surfaces, this represents the electrostatic potential of the surface of the given conductor.  Even when the actual value of the potential is not necessarily known, such as in the case of the isolated conductor, the condition of equipotential surfaces still holds (i.e. the potential is constant across the entire surface).  As will be shown later, the case is slightly altered when dealing with dielectric materials, though the mathematical foundations are unchanged.

In solving for the distribution of charges, the potential is therefore evaluated at the surface boundaries.  One often resorts to numerical methods involving discretization of the surface in order to solve for the charge distribution and hence determine the potential at any point in space.  The discretization is often carried to the point where it is possible to treat the surface charge density for a given element as uniform over the sub-element surface $dS'$.  In this case, Eq.~\ref{indirectBEMFinalForm} can be re-written in matrix-like notation:

\begin{equation}
U_i = \sum_{j=1}^N I_{ij} \cdot \sigma_j,
\label{eq:potential}
\end{equation}

\noindent where $U_i$ is the potential of the sub-element $i$, $\sigma_j$ is the charge density of sub-element $j$ of $N$ total elements, and $I_{ij}$ is given by the expression:

\begin{equation}
I_{ij} = \frac{1}{4\pi\epsilon_0}\int_{\Delta S_j} \frac{dS_j}{|\vec{r}_i - \vec{r}_j|}. 
\end{equation}

The element $I_{ij}$ therefore represents Green's function correlating sub-element $i$ at $r_i$ to sub-element $j$ at $r_j$.  Inversion of Eq.~\ref{eq:potential} yields the relevant discrete charge distributions from which Equation~\ref{indirectBEMFinalForm} can be fully evaluated.  However, for geometries where extremely large number of elements exists (warranted either by the required accuracy, the extent of the surface involved, or both), numerical inversion of Eq.~\ref{eq:potential} can impose often severe requirements, particularly on convergence and memory.  As we shall see in Section~\ref{sec:RH}, the Robin Hood method allows for elegant solution to the inversion/memory problem.  Before we leave the section, we expand the formalism so as to also include linear dielectric and magnetic materials.

\subsection{Extension to Dielectrics and Magnetic Materials}

In considering the calculation of electrostatic potentials in the presence of dielectric media, one no longer can impose the condition that each surface can be treated as an equipotential.  Nevertheless, it is certainly possible to recast the problem in terms of a matrix equation where one solves for the induced charges along the boundaries of the system~\cite{Rao}.  We still consider the case that Laplace's equation on the potential still holds:

\begin{equation}
\nabla^2 U(\vec{x}) = 0.
\end{equation}

Let us take the boundary condition that at the insulator-insulator interface the displacement vector $\vec{\bf D}$ is continuous across the surface.  If the surface represents the boundary between two dielectric materials, then the boundary condition on the surface imposed by Maxwell's equations results in:

\begin{equation}
\epsilon_i^+ \vec{\bf E}_i^+ \cdot \hat{n}_i - \epsilon_i^- \vec{\bf E}_i^-\cdot \hat{n}_i = 0,
\end{equation} 

\noindent where $\epsilon^{\pm}_i$ is the permittivity above and below the surface of the subelement $i$, $\vec{\bf E}_i^\pm$ is the value of the electric field at sub-element $i$, and $\hat{n}_i$ is the surface normal vector at the insulator-insulator interface.  The boundary condition can be re-expressed also in terms of the gradient of the potential:

\begin{equation}
\epsilon_i^+  (\hat{n}_i \cdot \vec{\nabla}) U_i^+ - \epsilon_i^-  (\hat{n}_i \cdot \vec{\nabla}) U_i^- = 0.
\end{equation} 

Here, the expression $(\hat{n}_i \cdot \vec{\nabla})$ represents the gradient oriented along the direction normal to the surface $i$.  In order to evaluate the electric field at a selected point, it is convenient once again to express everything in terms of their integral forms.  In this case, we can take advantage of the kernel evaluation of Green's function:

\begin{equation}
(\hat{n}_i \cdot \vec{\nabla}) G(\vec{\bf r}_i,\vec{\bf r}_j) = \frac{-1}{4\pi \epsilon_0} \frac{(\vec{\bf r}_i - \vec{\bf r}_j)\cdot \hat{n}_i}{|\vec{r}_i - \vec{r}_j|^3}.
\end{equation}

Hence, the electric field can be expressed as the following sum over finite elements:

\begin{equation}
\vec{\bf E}^{\pm}_i \cdot \hat{n_i} = \sum_{j=1, j\neq i}^{N} \frac{\sigma_j}{4\pi\epsilon_0}\int_{\Delta S_j} \frac{(\vec{\bf r}_i - \vec{\bf r}_j) \cdot \hat{n}_i}{|\vec{r}_i - \vec{r}_j|^3} dS_j \pm \hat{n}_i \cdot \hat{n}_i  \frac{\sigma_i}{2 \epsilon_0}.
\label{eq:EField}
\end{equation}

With some manipulation of the above equation, we can re-express it in matrix notation, just as we did for the scalar potential:

\begin{equation}
\Psi_i = \sum_{j=1}^{N} \eta_{ij} \sigma_j,
\end{equation}

\noindent where for $i\neq j$ we have

\begin{equation}
\eta_{ij} = \frac{1}{4\pi\epsilon_0}\int_{\Delta S_j} \frac{(\vec{\bf r}_i - \vec{\bf r}_j)}{|\vec{r}_i - \vec{r}_j|^3} \cdot \hat{n}_i dS_j,
\end{equation}

\noindent and for $i = j$ we have

\begin{equation}
\eta_{ii} = \frac{1}{4\pi\epsilon_0}\frac{2\pi (\epsilon^+_i + \epsilon^-_i)}{\epsilon^+_i - \epsilon^-_i}
\end{equation}

In the case of dielectrics, the boundary condition forces the constraint vector, $\Psi_i$, to be zero for all elements. The matrix equation can easily be extended to include a mixture of dielectric and conducting surfaces, as long as one calculates all contributions to the electric field and electric potential when evaluating the appropriate matrix elements.  Formulated in this manner, and taking advantage of the linearity of electric fields, one can impose the same algorithm to solve conductors and insulators.

The applicability of the Robin Hood method to both Neumann and Dirichlet boundary conditions implies that one can extend the method to the calculation of magnetic fields, even in the presence of materials with magnetic permeability~\cite{Garcia}.   For the case of magnetostatics in conductive media (no magnetic materials), one wishes to compute the magnetic field $\vec{B}(\vec{x})$ due to specified current configuration.  It is best to describe the system in terms of the magnetic vector potential, $\vec{\bf A}$, where $\vec{\bf B} = \vec{\nabla} \times \vec{\bf A}$.  If we decide to work in the Coulomb gauge

\begin{equation}
\vec{\nabla} \cdot \vec{\bf A} \equiv 0,
\end{equation}

\noindent then the vector potential can be written in a very similar manner as in electrostatics:

\begin{equation}
\vec{\bf A}(\vec{x}) = \frac{\mu_0}{4 \pi} \int_{\Gamma_\Omega} \frac{\vec{\bf K}(\vec{x}')}{|\vec{x}-\vec{x}'|}dS',
\end{equation}

\noindent where $\vec{\bf K}$ is the surface current density and $\mu_0$ is the permeability of free space.  If the current density is zero within the volume considered, then it is possible to express the magnetic field in terms of a magnetic {\em scalar} potential.  In such cases, one often has the knowledge of the current being distributed to the system, hence, performing a matrix inversion to solve for the free currents is not necessary.  However, once magnetic materials are introduced into the system, one no longer has knowledge of the induced currents, and one therefore returns to the same problem as posed in the case of dielectric materials.  Fortunately, we use an approach similar to that employed for dielectric materials to solve for such cases.  

Let us define the magnetic field above and below the boundary of some magnetic material as $\vec{\bf B}_i^{\pm}$.  Within the Coulomb gauge, the magnetic field can be expressed in terms of the vector potential $\vec{\bf A}$:

\begin{equation}
\vec{\bf B}_i^{\pm} = \sum_{j=1, j \neq i}^N (\nabla \times \vec{\bf A}_j(\vec{r}_i)) \pm \frac{\mu_0}{2} (\vec{\bf K}_i \times \hat{n}).
\label{eq:B}
\end{equation}

\noindent Here, $\vec{\bf K}_i$ is the surface current density due to element $i$, and $\vec{\bf A_j}(\vec{r}_i)$ is the magnetic vector potential at the point $\vec{r}_i$ due to all the other neighboring elements.   In the absence of free currents, the boundary condition on the sub-element $i$ above and below the surface is given by

\begin{equation}
\hat{n}_i \times (\frac{1}{\mu_i^+} \vec{\bf B}_i^+ - \frac{1}{\mu_i^-} \vec{\bf B}_i^-) = 0.
\label{eq:Bbound}
\end{equation}

Substituting Eq.~\ref{eq:B} into Eq.~\ref{eq:Bbound} yields the condition:

\begin{equation}
\hat{n}_i \times (\frac{1}{\mu_i^+} - \frac{1}{\mu_i^-})\sum_{j=1}^N (\nabla \times \vec{\bf A}_j(\vec{r}_i)) + (\frac{1}{\mu_i^+} + \frac{1}{\mu_i^-})\frac{\mu_0 \vec{\bf K}_i}{2}= 0.
\end{equation}

Dividing out the common term, one finds:

\begin{equation}
\sum_{j=1, j \neq i}^N \hat{n}_i \times (\nabla \times \vec{\bf A}_j(\vec{r}_i)) +  \frac{\mu_0(\mu_i^- + \mu_i^+)}{2(\mu_i^- - \mu_i^+)}\vec{\bf K}_i = 0.
\end{equation}

The magnetic vector potential term can be simplified further by realizing that the vector $\vec{\bf A}$ is continuous along the boundary

\begin{equation}
\hat{n} \times \vec{\bf B} = \hat{n} \times (\nabla \times \vec{\bf A}) = \nabla (\hat{n} \cdot \vec{\bf A}) - (\hat{n} \cdot \nabla) \vec{\bf A}.
\end{equation}

%
%
%
%
In order to simplify the constraints on each of the components of the current $\vec{\bf K}$, let us consider both the vector normal to the surface $S$, $\hat{n_i}$, as well as the two independent components tangential to the surface, $\hat{t}_i^l$, where $l=1,2$.  The boundary conditions impose different conditions on the tangential and normal components of $\vec{\bf K}$.  Specifically, the tangential boundary conditions can now be written as:

\begin{eqnarray}
\sum_{j=1, j \neq i}^N  \frac{-\mu_0}{4\pi}\int_{\Delta S_j} dS_j \frac{(\vec{\bf r}_i - \vec{\bf r}_j)}{|\vec{\bf r}_i - \vec{\bf r}_j|^3} \cdot (\hat{t}_i^l (\hat{n}_i \cdot \vec{\bf K}_j ) - \hat{n}_i ( \hat{t}^l_i\cdot \vec{\bf K}_j))\\
  ~+ \frac{\mu_0 (\mu_i^- + \mu_i^+)}{2(\mu_i^- - \mu_i^+)}(\hat{t}_i^l \cdot \vec{\bf K_i}) = 0.
\end{eqnarray}

If once again we assume the surface current does not change across the integration, we can express this as a matrix whose elements are given as follows

\begin{equation}
\Psi_i^l = \sum_{j=1}^{N} \vec{\chi}_{ij} \cdot (\hat{n}_i  (\hat{t}_i^l \cdot \vec{\bf K}_j) - \hat{t}^l_i (\hat{n}_i  \cdot \vec{\bf K}_j )),
\end{equation}

\noindent where for $i \neq j$ we have

\begin{equation}
\vec{\chi}_{ij} = \frac{\mu_0}{4\pi} \int_{\Delta S_j} \frac{(\vec{\bf r}_i - \vec{\bf r}_j)}{|\vec{r}_i - \vec{r}_j|^3} dS_j,
\end{equation}

\noindent and for $i=j$ we have

\begin{equation}
\vec{\chi}_{ii} =   (\frac{\mu_0}{4\pi})\frac{2 \pi (\mu_i^- + \mu_i^+)}{\mu_i^+ - \mu_i^-}  \hat{n}_i.
\end{equation}

%
%
%
%

Since at the surface $i$ the current is constrained to flow along the surface, there is no normal component $\hat{n}_i \cdot \vec{\bf K}_i$. 

These conditions completely define induced surface currents and surface charges from all elements. For linear materials, these equations can be solved to obtain the values of $\sigma$ and $\vec{\bf K}$ across all surfaces by means of once again inverting an $N \times N$ matrix.  Note that we have restricted ourselves to the case where there are no external source charges or currents in the volume.  Had this assumption been relaxed, the boundary vector $\Psi_i$ would take on values other than zero.

\section{Technique and Implementation}
\label{sec:Technique}

\subsection{The Robin Hood Method}
\label{sec:RH}

Let us consider the case where an isolated conducting sphere comes in close proximity to a charge (or collection of charges) some distance away.  Although the total charge on the sphere is zero, the charges rearrange themselves so as to ensure that the electric potential everywhere on the conduct is constant, thereby ensuring Gauss' law is satisfied.  If there is an imbalance in the potential, charges move and swap until the proper potential at the surface is reached.  Nature, of course, accomplishes this nearly instantaneously but the rules that guide the distribution of charges can be simplified such that computer algorithms can be made to mimic the effect.  

The Robin Hood method optimizes this natural strategy to conform on how computer algorithms are designed, while maintaining the simplicity of the rules illustrated briefly above.  We will discuss the details of the Robin Hood implementation in the next section.

\subsection{Implementation}

Consider once again our example of the isolated conducting sphere in close proximity of a single charge of value $+q$ located at some distance away.  As first step, which embeds the approximation scheme used in this approach, we subdivide the geometry of our large sphere into $N$ discrete triangles and assume a constant charge density across the surface of each triangle.  The discretization determines the level of accuracy achievable by the model.  Since the charge distribution is unknown, we randomize the individual charge distributions for each triangle.  If the object had some finite charge, the constraint would be imposed that the sum of all charge add up to the assigned charge. The potential is then calculated from this initial assignment of charge distributions, in accordance with Equation~\ref{eq:potential}.


The bulk of the computing time is spent in calculating the matrix coefficients $I_{ij}$, which essentially reduces to computing the potential from a triangle at some arbitrary point $\vec{x}_i$.  We have the option of employing a variety of techniques in computing this potential.  In cases where accuracy precedes timing needs, we have the option of computing the potential from a triangle analytically.  Such a computation is a bit more time consuming, as the evaluation has terms that depend on $\sinh^{-1}(x)$ and $\tanh^{-1}(x)$ functions.  Alternatively, one can use self-refining mesh and simply compute the multipole expansion up to the quadrupole term to obtain the value of the potential at the given point. The criterion for accuracy using this approach is expressed as a ratio (let us denote it by $\nu$) of the distance $|\vec{x}_i -  \vec{x}_ j|$ and the typical size of the right-angled triangle which is characterized by the length of the larger leg of the triangle. When the ratio $\nu$ is larger than some pre-determined cutoff (in our case, 27), the monopole term alone is enough to obtain the value of the potential at point $\vec{x}_j$ with 4 digit accuracy. If the ratio $\nu$ is smaller that 27 but larger than 5.7 then monopole and quadrupole term together yield a 4 digit accuracy. If the ratio $\nu$ is smaller than 5.7 then the initial triangle is to be divided into four new triangles. In that case, the potential becomes the sum of 4 contributions from the new triangles, which are considered homogeneously charged bearing the same charge density as the initial triangle.  A full list of the expressions used in the triangle potential evaluation can be found in the Appendix.

Once the potential is evaluated at each boundary element in the problem, the algorithm then performs a search for the two elements that differ most from the equipotential surface.  Assume that two of the worst
triangles are electrodes $m$ and $n$.  The equipotential condition imposes that these two elements have the same potential. Therefore, a small amount of charge is moved from one electrode to the other such that the equipotential condition is met; that is, $U_m' = U_n'$. The new potential due to this redistribution is therefore given by the expression 

\begin{eqnarray}
U_m' = U_m + I_{mm} \cdot \delta \sigma_m + I_{mn}  \cdot \delta \sigma_n, \\
U_n' = U_n + I_{nn}  \cdot \delta \sigma_n + I_{nm}  \cdot \delta \sigma_m.
\end{eqnarray}


\noindent where $U_{m,n}$ is the potential at the two selected points. For the case of two sub-elements, there is an exact solution to the amount of charge that needs to be moved from one element to the other.

\begin{eqnarray}
\label{eq:dq}
\delta \sigma_m = \frac{A_n (U_n - U_m)}{A_n (I_{mm} - I_{nm}) + A_m (I_{nn}- I_{mn})}, \\
\delta \sigma_n = \frac{A_m (U_m - U_n)}{A_n (I_{mm} - I_{nm}) + A_m (I_{nn}-I_{mn})},
\end{eqnarray}

\noindent where $\delta \sigma_{m,n}$ is the charge density exchanged and $A_{m,n}$ is the area of the triangle sub-element.  Note that the sum of the two charges is zero; i.e. $\delta q_{\rm total} = \delta \sigma_m \cdot A_m + \delta \sigma_n \cdot A_n = 0$.  Though we have used the example of an isolated sphere to illustrate the essential elements of the algorithm, we can certainly extend the case for a conductor held at a fixed potential.  When the sphere is held at a fixed potential, $U_0$, the shift of charges between the two electrodes is altered to the following expression:

\begin{eqnarray}
\delta \sigma_m = \frac{(U_0 - U_m)I_{nn} - (U_0-U_n)I_{mn}}{I_{mm} I_{nn} -I_{mn}I_{nm}}, \\
\delta \sigma_n = \frac{(U_0 - U_n)I_{mm} - (U_0-U_m)I_{nm}}{I_{mm} I_{nn} -I_{mn}I_{nm}}.
\label{eq:charge_exchange}
\end{eqnarray}

Having solved for the shift in charge, the potential for every sub-element is subsequently updated to reflect the new charge configuration.  The method is said to converge once the maximum and minimum values of the potential on an individual electrode is below some user-defined value.  In all examples used to date, convergence appears to be scale-independent, i.e. it continues to fall exponentially until machine precision is achieved.  As can be seen from the above description, the algorithm is extremely simple in implementation, yet encodes the basic natural properties imposed by Gauss' law.  

\subsection{Some Generalizations}

So far we have restricted ourselves to the case where just two electrodes are involved in the charge exchange.  We can generalize the method to the case of $M$ electrodes exchanging charge, where $1 \le M \le N$.  For the case where the electrodes are held at fixed potential, the charge exchange solution is equivalent to the inversion of a $M \times M$ matrix:

\begin{equation}
U_i' = U_i + \sum_{j = 1}^{M} I_{ij} \delta \sigma_j,
\end{equation}

\noindent where $U_i'$ is the adjusted potential after a small amount of charge $\delta q_j = \Delta S_j \cdot \delta \sigma_j$ is exchanged between $M$ selected electrodes. Both indexes $i$ and $j$ range from 1 to $M$. Note that for the case that $M = 2$ and the target potential is known (i.e. $U_i' = U_0$ for all selected electrodes), one reproduces Eq.~\ref{eq:charge_exchange}.  

The same generalization can be made for the insulated charge case.  Here, one must explicitly introduce the conservation of charge as part of the matrix to be solved.   In general, one can impose the conservation of charge as:

\begin{equation}
\sum_{j=1}^{M} \delta q_j = 0.
\end{equation}

As for the condition imposed by the equipotential surface, all the potentials at each altered electrode are identical.  This condition can be enforced via the recursive relation:

\begin{equation}
\sum_{j=1}^{M}(I_{i+1,j} - I_{i,j}) \delta \sigma_j = U_{i}-U_{i+1}
\end{equation}

The index $i$ ranges from 1 to $M-1$.  The combination of the two constraints produces an $M \times M$ matrix equation which can be inverted to solve for $\sigma_j$.  Note that for the isolated conductor we require that $M > 1$.  In the case where $M=2$, one reproduces Eq.~\ref{eq:dq}.  It is interesting to study the properties of these solutions for various limits on the value of $M$.  For the case $M \rightarrow 1$, the Robin Hood method becomes a variant of the Gauss-Seidel method, albeit it is implemented with a much lighter memory footprint.  In the limit $M \rightarrow N$, it approaches the original matrix inversion problem.  In general, the convergence behavior is best for $M \rightarrow 1$.  

There is one last generalization that can be useful in practical applications.  In most study cases, one is not just interested in a single configuration, but in the behavior of the system under different boundary conditions.  As long as the geometry remains fixed, it is possible to solve the charge configuration to allow any arbitrary combination of voltages.

Consider the case where one can break up the boundary into different groups of electrodes (each group corresponds to a subspace of $\Gamma_\Omega$), each of which is held at a distinct potential that is common to all the elements in that subgroup.  It is now possible to solve for the charge configuration for the case where all the elements in a given subgroup are held at a common potential, $\Phi_0$, while all the other potentials are held at ground; that is:

\begin{eqnarray}
U^k_i = \sum_j I_{ij} \sigma^k_j,
\end{eqnarray}

\noindent where $k$ denotes the specific group, $U_i$ is zero everywhere except for the electrodes in group $k$, and $\sigma^k$ represents the charge solution for the group.  This procedure is repeated for each and every group in the set.  Because the electric potential everywhere in space depends on the linear superposition of the charges, we can now construct the electric potential due to any arbitrary setting of the potentials, $\Phi_i$, on each of these groups: 

\begin{equation}
\sigma_i^{\rm sup} = \sum_{k = 1}^{N_g} \alpha_k \sigma_i^k,
\end{equation}

\noindent where $N_g$ is the total number of electrode groups in the configuration and $\alpha_i$ is a scalar defined as

\begin{equation}
\alpha_i = \Phi_i / \Phi_0.
\end{equation}

Although the computation time obviously increases approximately by $N_g$, further computation is no longer necessary even if the potentials on the electrodes are altered.

\subsection{Convergence Properties}

The Robin Hood approach is expected to converge as long as the following three conditions are satisfied:

\begin{enumerate}
\item The system is linear.
\item The matrix is diagonally dominant.
\item The charge configuration corresponding to the boundary conditions is unique.
\end{enumerate}

Conditions (i) and (ii) are satisfied by the properties of Maxwell's equations and Green's function.  Indeed, since the potential falls like $\frac{1}{|\vec{x}_i - \vec{x}_j|}$, the self-term diagonal element in the matrix is always greater than its neighbors, and so convergence is guaranteed.  One could argue that in expression~\ref{eq:dq} the denominator $I_{mm} + I_{nn} - I_{mn} - I_{nm}$ could become zero thus causing the Robin Hood approach to fail. However, for a homogeneously charged triangle the potential has the largest value at the barycenter of the triangle, which implies that for any other element $n$ the relation $I_{nn} > I_{nm}$ always holds. In particular, if it happens that $I_{nn}$ is equal to $I_{nm}$, it implies that the two different triangles share the same barycenter, which means that they are either intersecting or completely overlapping. Neither of these situations should be allowed because in that case the whole electrostatic problem becomes under defined thus allowing for infinitely many solutions. Therefore, in well defined geometries, condition (iii), and by inference condition (ii), are always satisfied.

\subsection{Advantages}

Despite its extremely simple algorithm, the Robin Hood method offers a number of concrete advantages over other matrix inversion methods.  The main advantage of the technique is its memory footprint.  Under the Robin Hood approach, there is no need to store individual matrix elements during the computation.  Rather, only the potential at each element is stored, while the matrix elements are simply re-calculated when updating the potential due to the charge exchange.  As such, the memory requirements grow linearly with $N$, instead of quadratically.  This feature allows for solving the charge configuration even for extremely segmented geometric configurations previously avoided.

The other advantage inherent to the technique is that it is extremely parallel in nature.  The computation of potential can be distributed to any number of CPUs, since each computation is essentially isolated.  In fact, with the exception of the charge exchange calculation itself, the entire sequence can be passed to a cluster of machines to facilitate computation.  For this paper, the calculations are typically performed on clusters using the Message-Passing Interface (MPI) protocol with great success in reducing real calculation time.  The Robin Hood method also has recently been extended to function on Graphical Processing Units using the OpenCL standard~\cite{OpenCL}, and has been verified on both ATI and NVIDIA graphics cards.

One last advantage that Robin Hood provides is that the computation time grows quadratically with $N$, while using direct solvers the growth is cubic.

\section{Case Studies}
\label{sec:systems}

\subsection{Verification Examples}

The Robin Hood algorithm has already been benchmarked in previous publications against standard examples used in electrostatics.  Most notably, the Robin Hood method has been used to calculate the capacitance of a cube.  The largest cube calculated was represented by 202,800 triangles, obtaining a capacitance value of $0.66067786 \pm 8 \times 10^{-8}$ in units of $1/4\pi\epsilon_0$.  The calculation represents one of the most accurate values of the capacitance of the cube calculated to date~\cite{RH1,RH2}.  The algorithm is not limited in terms of the number of triangles employed and, when used in combination with parallel processing, can be remarkably fast and efficient.

\begin{figure*}[htb]
\begin{center}
{
\begin{tabular}{c c}
\includegraphics[width=0.5\linewidth,clip=true]{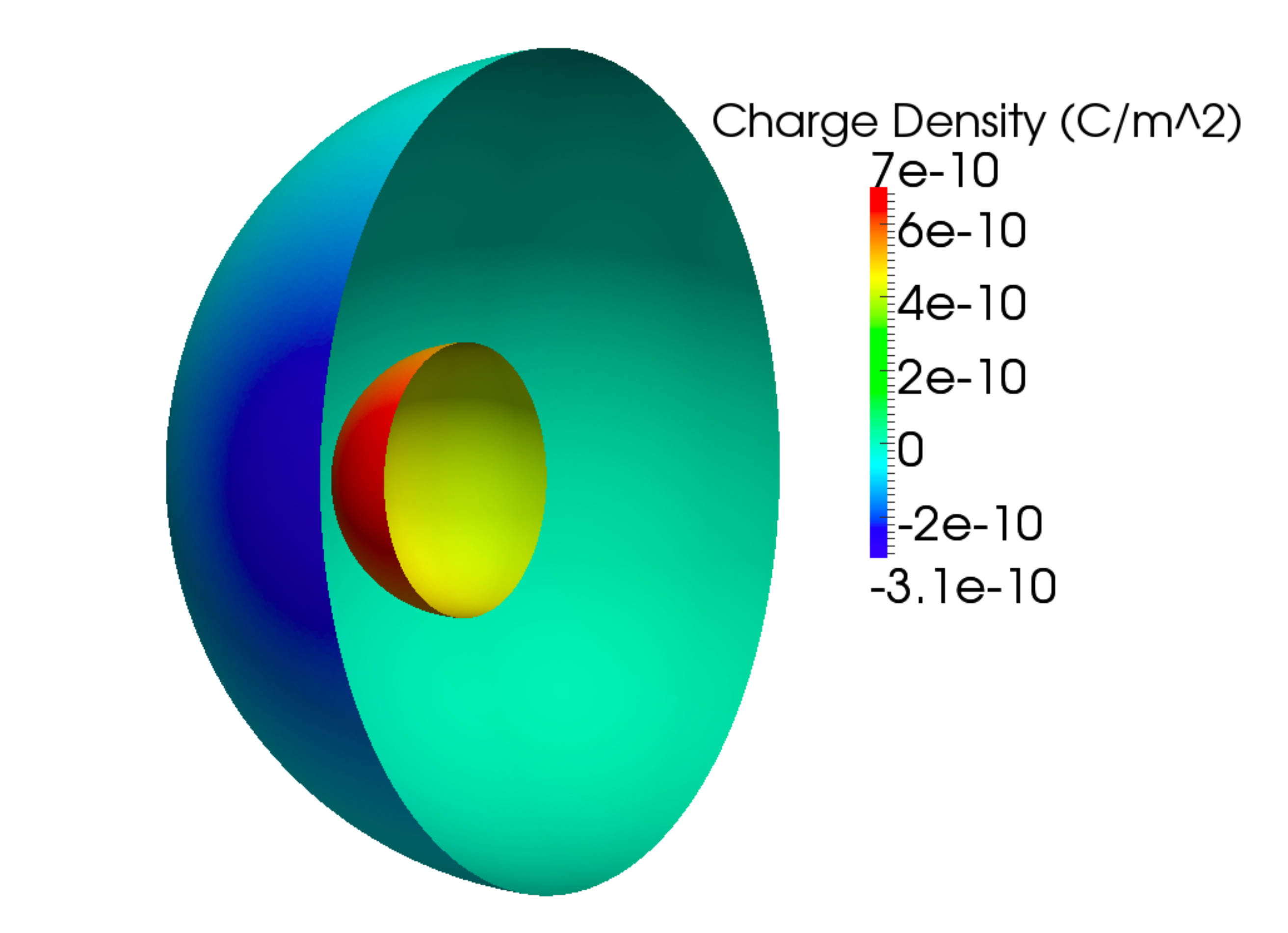} &
\includegraphics[width=0.5\linewidth,clip=true]{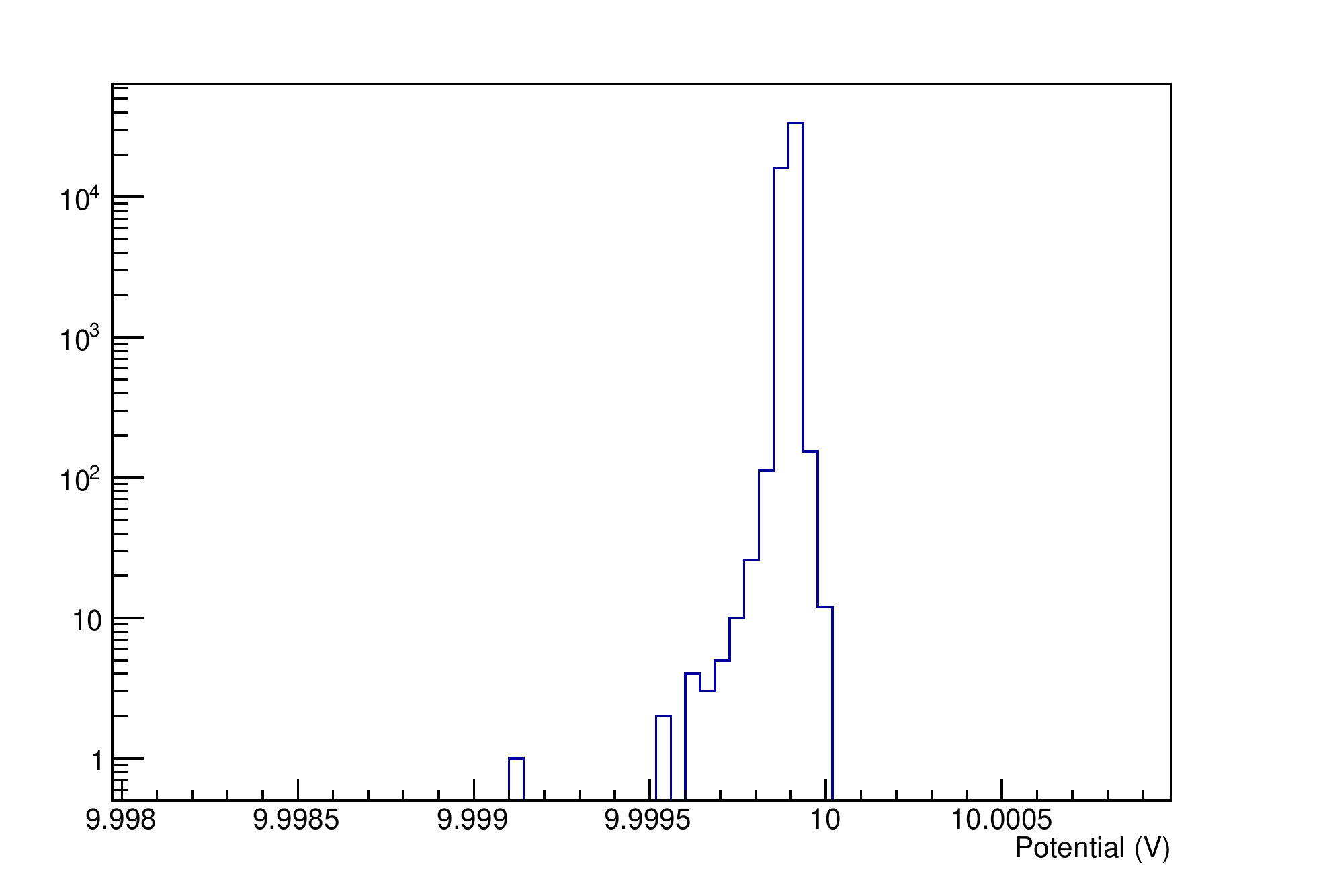} \\
\end{tabular}
}
\end{center}
\caption{Left: Two nested conducting spheres, one held at ground potential while the other held at a fixed potential of 10 V. Right: Distribution of potential value for (solved) nested sphere geometry on left.  The potential distribution is sampled throughout the volume of the innermost sphere.}
\label{fig:nestedspheres}
\end{figure*}

The capacitance is a poor metric for testing the accuracy of the method, in part because it is an integral measure of the system variation.  A more compelling comparison is one that tests certain fundamental principles, such as Gauss' law.  We consider the case where we have a set of nested metal spheres, where the inner sphere is held at a fixed potential (10 V) while the outer shell is held at ground.  Gauss' law predicts that the electric potential inside the innermost sphere is fixed, regardless of whether the inner sphere is concentric with the outer sphere or not.  For our study, the sphere was subdivided into 20,000 triangles using GMSH's Delaunay algorithm~\cite{GMSH}. The accuracy of the calculation is determined by the metric

\begin{equation}
\bar{\epsilon} = \sum_i^{N} |U(\vec{x}_i) - U_{\rm target}|.
\end{equation}

The quantity $\bar{\epsilon}$ is sampled throughout the volume inside the innermost sphere. Results for our test can be seen in Figure~\ref{fig:nestedspheres}.   For the example cited above, the average voltage within the sphere was within $\pm 15 \mu$V of its expectation value.

\subsection{M- and N- Dependence Studies}

\begin{figure*}[htb]
\begin{center}
{
\includegraphics[width=0.9\linewidth,clip=true]{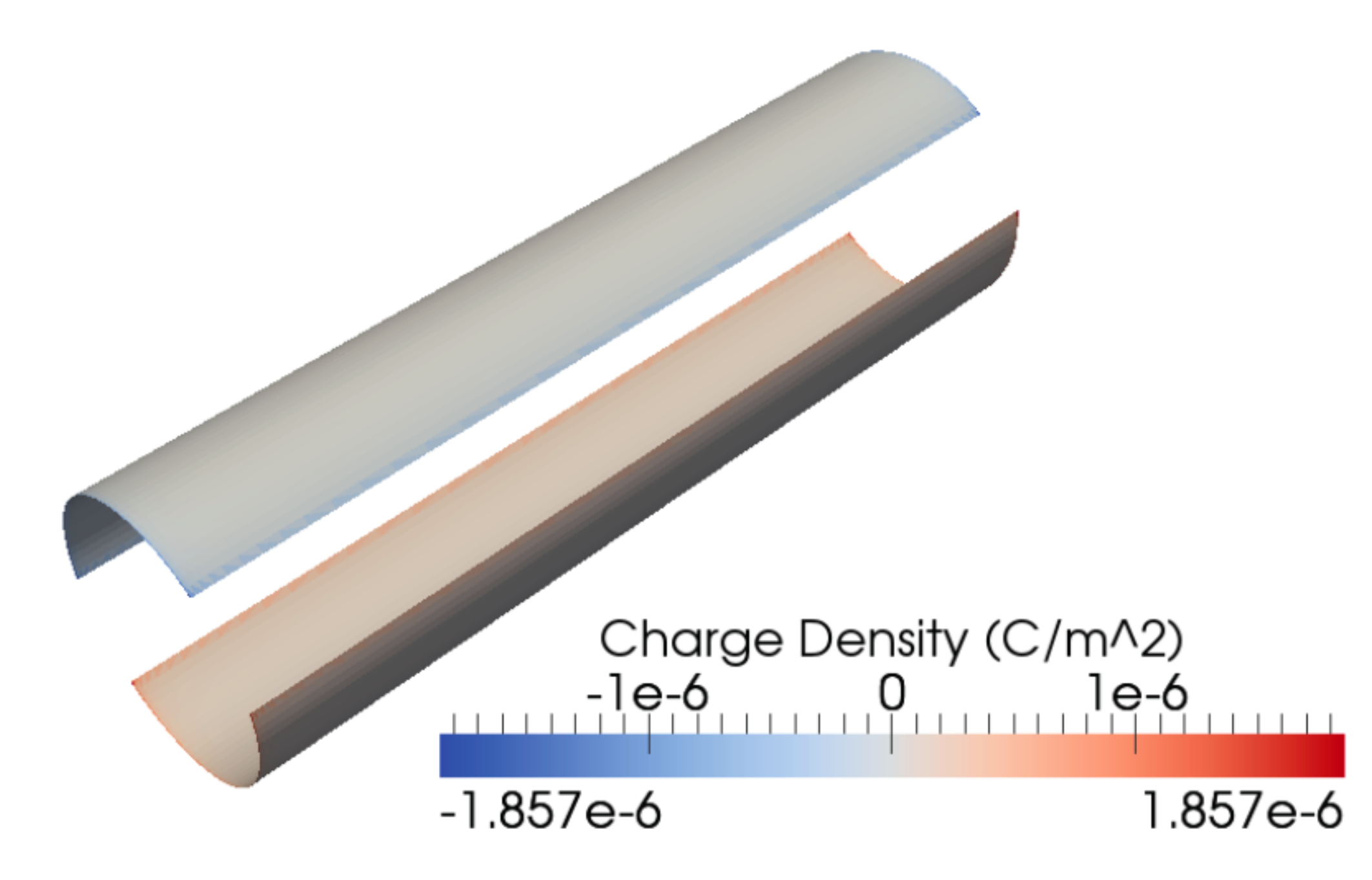}
}
\end{center}
\caption{A model of the dipole configuration used for M- and N- dependence studies.  The color indicates the surface charge density distribution of the object.}
\label{fig:Dipole}
\end{figure*}

In order to test some of the fundamental properties of the Robin Hood method, we made electric
field computations of a simple dipole electrode system. This consists of two electrodes having
opposite potentials (1000 V and -1000 V), each of them having a half-cylindrical shape, with some
gap between them (see Figure~\ref{fig:Dipole}). We discretized the surface of these electrodes with $N$ independent triangle surfaces. All iterations were started with zero charge density on all triangles.

We investigated the M-dependence of the potential accuracy and of the iteration number,
with a discretization of $N=3600$. Table \ref{TableM}
contains results of the dipole charge solution computed with 6 different $M$ values. We study both the relative potential accuracy $A_{\rm rel}$ values achieved by the Robin Hood method with fixed $M\cdot n$ product and with fixed relative accuracy. The relative accuracy $A_{\rm rel}$ denotes the maximal deviation between the input potential and the simulated potential, over all triangle subelements.  Since the computation time $t$ of the Robin Hood method is proportional to $M\cdot n$, the computation time is roughly constant.
As Table  \ref{TableM} shows, we found the best accuracy with $M=1$ (the Gauss-Seidel limit of the Robin Hood method).
For a fixed relative accuracy, it appears that $M=1$ is also the best choice for minimizing the computation time.

\begin{figure*}[htb]
\begin{center}
{
\includegraphics[width=0.9\linewidth,clip=true]{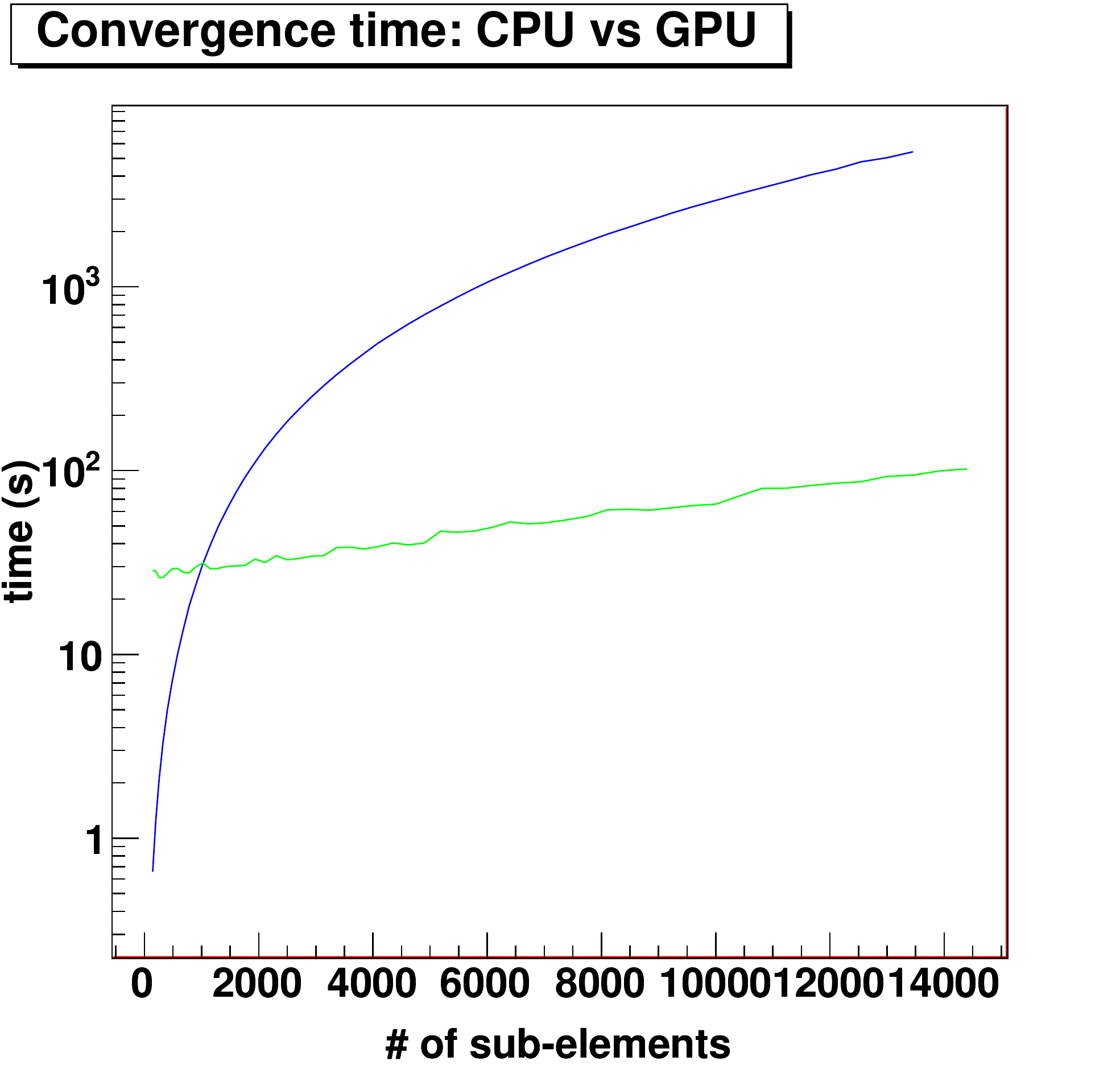}
}
\end{center}
\caption{A plot of the time dependence versus number of electrodes for the dipole configuration.  The plot shows the computation time for CPU processing (blue) and GPU processing (green) versus number of sub-elements.  In all cases, the required relative accuracy of the potential at the surface of each triangle electrode is set to be $10^{-8}$. }
\label{fig:Ndep}
\end{figure*}

Table \ref{Tablen} presents the number of iterations $n$ for various accuracy levels, using the same
discretization as above, and $M=1$. It seems that $n$ is approximately a linear  function
of $\log A_{\rm rel}$.  Figure~\ref{fig:Ndep} shows the time dependence of the RH solver as a function of the number of sub-elements once the relative accuracy is fixed. This time evolution is nearly quadratic in $N$.  These results are also shown in Table~\ref{TableN}. As a testament to the vast improvement to be gained from GPU processing, Figure~\ref{fig:Ndep} also shows the time evolution for the dipole configuration as solved with GPU implementation. 

We have repeated these simulations with several other electrode systems: unit square and unit cube, various electrodes containing both rectangles and wires as subelements
etc. In all cases, we obtained the same results as above:

\begin{itemize}
 \item The Robin Hood method is fastest with $M=1$ (Gauss-Seidel limit).
 \item The iteration number increases logarithmically with the relative potential accuracy.
\item The iteration number increases linearly and the computation time quadratically
with the number of the discretization subelements (for the CPU implementation).
\end{itemize}
 
\begin{table*}[!h]
\begin{center}
\vspace{2mm}
\begin{tabular}{|l|c|c|c|c|c|c|} \hline
 $M$ & 1 & 2 & 3 & 4 & 6 & 10  \\ \hline \hline
$A_{\rm rel}$  & $7.23 \cdot 10^{-12}$  &  $4.5 \cdot 10^{-11}$ & $8.3 \cdot 10^{-11}$ & $1.6 \cdot 10^{-10}$ &$3.9 \cdot 10^{-10}$  &$3.2 \cdot 10^{-9}$   \\
\hline
$M  \cdot n$  & 20999 & 22580 & 23547 & 24532 & 26796  & 29750 \\
\hline
\end{tabular}
\caption{Robin Hood results with a dipole electrode ($N=3600$), for 6 various
M values. Second row: relative accuracy  $A_{\rm rel}$ for fixed $M\cdot n=30000$ value
($n$: number of iterations).
Third row: $M\cdot n$ for
fixed relative accuracy $A_{\rm rel}=10^{-8}$.
\label{TableM}} 
\end{center}
\end{table*}

\begin{table*}[!h]
\begin{center}
\vspace{2mm}
\begin{tabular}{|l|c|c|c|c|c|} \hline
 $A_{\rm rel}$  & $10^{-2}$  &  $10^{-4}$ & $10^{-6}$ & $10^{-8}$ &$10^{-10}$    \\
\hline \hline
$n$  & 3425   &9108  &15047  & 20999 &  26973 \\
\hline
\end{tabular}
\caption{Robin Hood results with a dipole electrode ($N=3600$): 
number of iterations $n$ for various relative accuracy values $A_{\rm rel}$.
\label{Tablen}} 
\end{center}
\end{table*}

\begin{table*}[!h]
\begin{center}
\vspace{2mm}
\begin{tabular}{|c|c|c|c|c|c|} \hline
 $N$ & 900 & 1444 & 3600 & 7056 & 13456  \\ \hline \hline
n  &  5279 &  8496 & 20999  & 40859  & 77030   \\
\hline
$n/N$  & 5.87 & 5.88 & 5.83 & 5.79 & 5.72  \\ \hline
$t/N^2$  & $3.0\cdot 10^{-5}$ & $3.0\cdot 10^{-5}$ & $2.9 \cdot 10^{-5}$ & $2.9\cdot 10^{-5}$ &  $3.0\cdot 10^{-5}$ \\ 
\hline
\end{tabular}
\caption{Robin Hood results with a dipole electrode for 5 various
discretizations, with fixed relative accuracy $A_{\rm rel}=10^{-8}$, as computed with CPU implementation.
Number of independent subelements: $N$, 
number of iterations: $n$, computation time (in seconds): t.
\label{TableN}} 
\end{center}
\end{table*}

Having verified the efficacy of the method and its optimal performance, we now choose two very distinct examples --a large-scale system and a micro-scale system -- to illustrate the technique's usefulness and versatility in real-world applications.

\subsection{Case 1: Large-Scale System:  the KATRIN Focal Plane Detector}

The KArlsruhe TRItium Neutrino experiment (KATRIN) combines an ultra-luminous molecular tritium source with an integrating high-resolution spectrometer to gain sensitivity to the absolute mass scale of neutrinos. The projected sensitivity of the experiment on the neutrino mass scale is 200 meV at 90\% C.L.~\cite{bib:KATRIN}.  A brief description of the overall experimental setup, currently under construction, is given below.

\begin{figure*}[htb]
\begin{center}
{
\includegraphics[width=0.9\linewidth,clip=true]{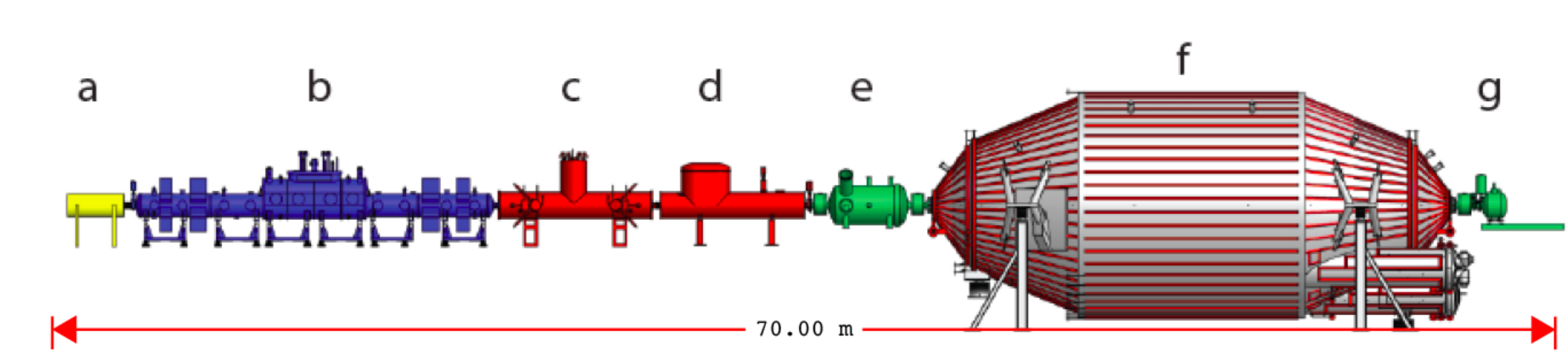}
}
\end{center}
\caption{Schematic overview of the KATRIN experimental setup: (a) rear section, (b) tritium source, (c) diffrential pumping section, (d) cryogenic pumping section, (e) pre-spectrometer, (f) main spectrometer, (g) detector system. The overall setup has a length of about 70 m.  The Robin Hood model is based on the detector system (g) only.}
\label{fig:KATRIN}
\end{figure*}

The KATRIN experiment is based on technology developed by the Mainz and Troitsk tritium beta decay experiments. These experiments used a MAC-E-Filter (Magnetic Adiabatic Collimation combined with an Electrostatic Filter)~\cite{bib:MACE}. The technique, which combines high luminosity with high energy resolution is as follows:  Low-pressure tritium gas is stored in a high magnetic field tube.  Beta-decay electrons from the gas, though emitted isotropically, are guided longitudinally along the magnetic field towards a large spectrometer, while neutral gas is returned to the source tube by pumps (see Fig.~\ref{fig:KATRIN}).  On their way into the center of the spectrometer, the electrons follow the magnetic field while its strength drops by many orders of magnitude. The magnetic gradient force transforms most of the cyclotron energy E$_{\perp}$ into longitudinal motion.  The resulting parallel beam of electrons runs against an electrostatic potential formed by cylindrical electrodes.  Electrons with enough energy to pass the electrostatic barrier are reaccelerated and collimated onto a detector, while lower energy electrons are reflected. Therefore, the spectrometer acts as an integrating high-energy pass filter, with which we measure the $\beta$ spectrum in an integrating mode.   More detailed descriptions of each component can also be found in the KATRIN Design Report~\cite{bib:KATRIN}.  

\begin{figure*}[htb]
\begin{center}
{
\begin{tabular}{c c}
\includegraphics[width=0.4\linewidth,clip=true]{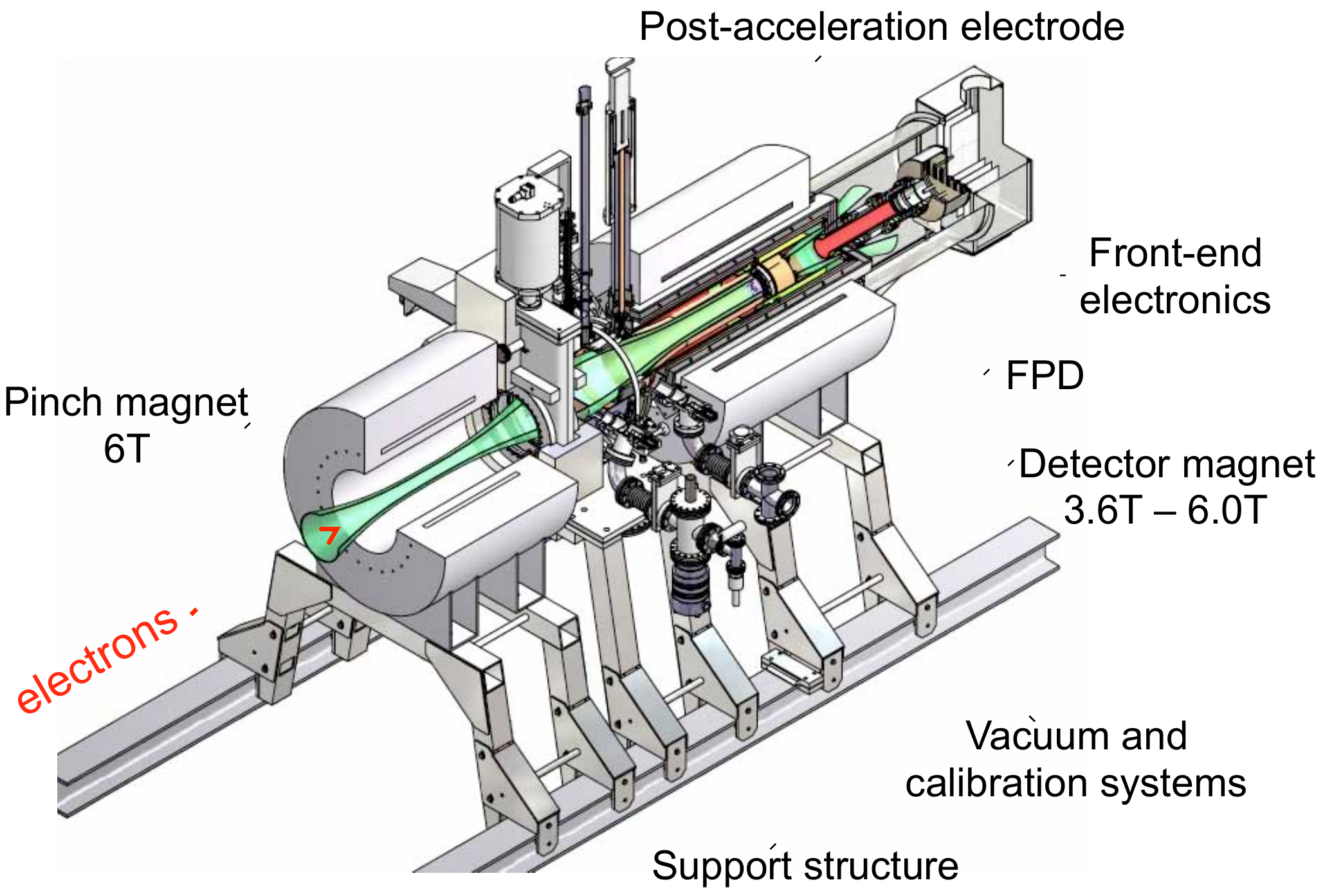} &
\includegraphics[width=0.4\linewidth,clip=true]{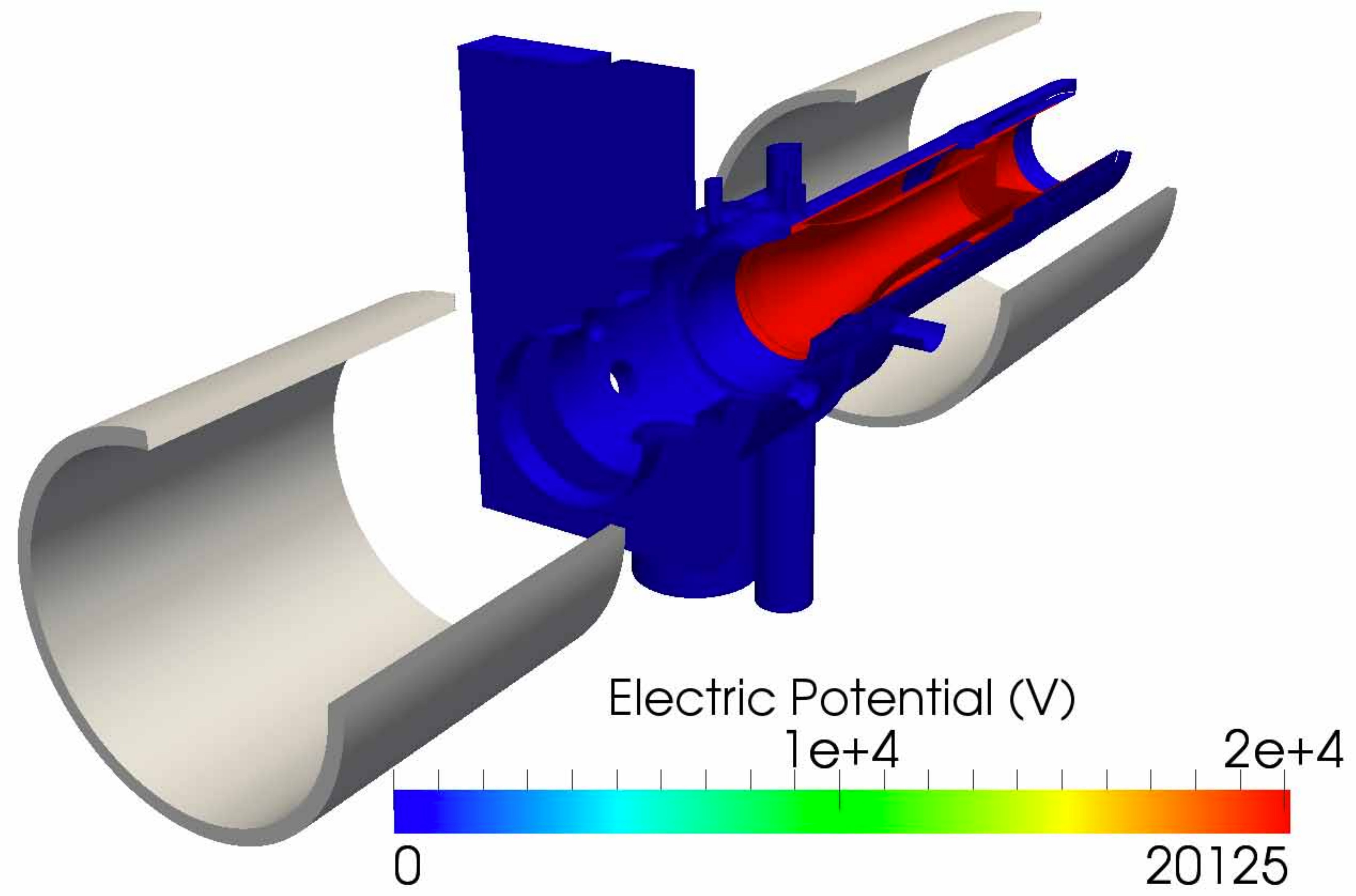} \\
\end{tabular}
}
\end{center}
\caption{Left: the primary components of the detector system.  Right:  The Robin Hood model of the detector region, solved for the potentials applied to the electrode surfaces.}
\label{fig:FPD}
\end{figure*}

In this study, we concentrate on the modeling of the focal plane detector (FPD), where the $\beta$-decay electrons are focused and detected.  An illustration of the FPD is shown in Figure~\ref{fig:FPD}.  Electrons enter left of the first (pinch) magnet and are magnetically collimated onto a silicon pin diode detector.  A horn-shaped electrode with a quartz insulator surrounds the magnetic flux lines in order to provide additional acceleration to electrons entering the system as a means to mitigate against low energy backgrounds.  Calibration and pumping ports can be seen at the outer edges of the vacuum tube, making the system's geometry highly asymmetric.  

The system's asymmetric geometry forces one to model the system using triangle discretization.  The experiment is sensitive to the presence of low level Penning traps (instances where magnetic field lines stretch across two separated electrodes, both at non-zero potential).  Such Penning traps are often a source of electrons and ions that contribute to the background of the experiment.  Finding such traps requires high level of geometric discretization and accuracy.  The memory requirement imposed by such high level discretization is often prohibitive for other electrostatic solvers.

The focal plane region was broken up into several electrode groups (each group is defined by whether it is a dielectric or held at a specified potential) and sub-sequentially divided into a total of 444,821 triangle sub-elements of varying scale.  Scaling is usually determined by one's proximity to intersecting structures or to sharp edges.  Calculation of the surface charge densities for this electrostatic configuration was carried out at the University of North Carolina computing grid, which consists of 520 blade servers, each with 2 quad-core 2.3 GHz Intel EM64T processors, 2x4M L2 cache (Model E5345/Clovertown), and 12 GB of memory.  The calculation was distributed using MPI protocol over approximately 100 CPUs.  A total of $\sim 1.5 \times 10^{6}$ iterations were carried out to each a minimum relative accuracy of $\epsilon \le 10^{-8}$ on all sub-elements.  The total computation time was approximately 16 CPU-months ($\sim 4$ days real time).

\begin{figure*}[htb]
\begin{center}
{
\includegraphics[width=0.9\linewidth,clip=true]{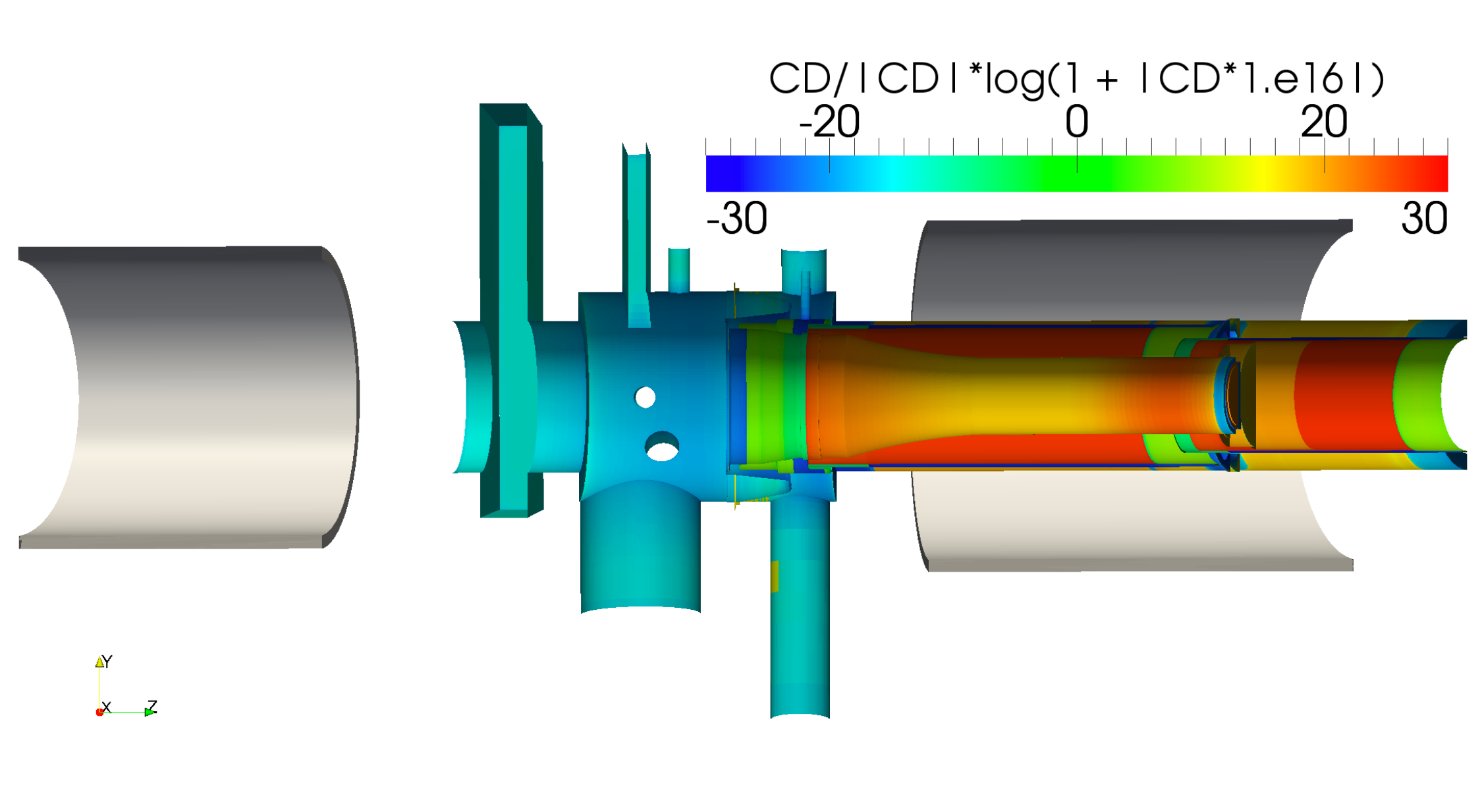}
}
\end{center}
\caption{Charge density distribution for section of detector region.}
\label{fig:DetectorChargeDensity}
\end{figure*}

\begin{figure}[htb]
\begin{center}
{
\includegraphics[width=0.9\linewidth,angle=270,clip=true]{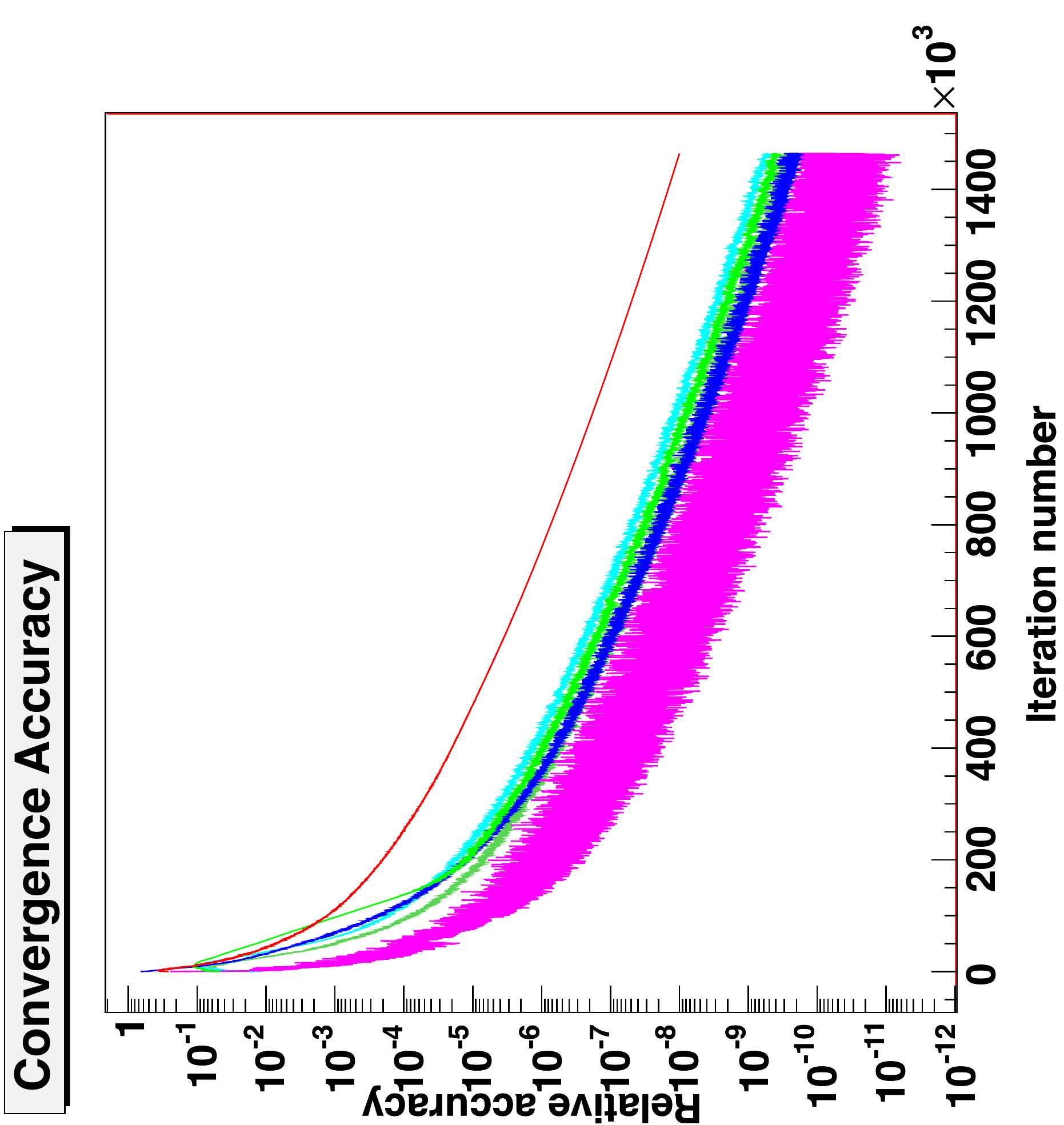}
}
\end{center}
\caption{Relative accuracy of detector electrode groups (represented as different curves) as a function of iteration.  The different curves represent different component groups which share the same target potential within the detector.  The convergence of the geometry scales exponentially with the number of iterations.  The calculation was performed until all electrode groups achieve a relative accuracy of $10^{-8}$.  }
\label{fig:convergence}
\end{figure}

Figure~\ref{fig:DetectorChargeDensity} shows the solved charge density distribution of the detector system after convergence.  Despite its high level of discretization, presence of dielectrics, and geometric complexity, the charge configuration is well defined throughout the volume, and its convergence is achieved in a reasonable time frame.  The convergence scale as a function of iteration is shown in Figure~\ref{fig:convergence}.  The charge configurations can be used to calculate various aspects of the electrostatic configuration, such as the presence of Penning traps and sources of potential vacuum discharge. The results of this study are the topic for an upcoming paper.  

\subsection{Case 2: Micro-Scale System: Nanostructuring at surface}

Field enhancement on nanostructures at the surface of electrodes is of great practical importance, particularly in the fields of nanoplasmonics \cite{plasmonics} and the development of ion emitters \cite{berkeley_guys}. Nanoplasmonics is a bit more complex, since in principle it includes the time-dependence of the electric fields\cite{plasmonics_retarded}, so we chose for this study to concentrate solely on ion emitters.

Ion emitters are designed to produce a relatively large current by ionizing atoms close to the surface of an electrode.  Field ionization occurs via electron tunneling on through a potential barrier between a neutral atom and an electrode maintained at a fixed voltage. The probability of tunneling depends of the strength of the electric field near the electrode and, as such, depends on both the potential difference applied to the electrode as well as the specific geometry of the electrode itself \cite{improved_Fowler_Nordheim_field_emission}. Due to the dependence of the electric field strength on the geometry, modeling of the properties of the ion emitter, such as field enhancement, is non-trivial.  In many applications, nanostructures are deliberately produced to maximize field-enhancement~\cite{berkeley_guys}.  This allows for simpler power requirements and often enables such devices to be much more portable~\cite{surface_state_enhancement,brain_signal_field}. 


The optimization of different geometries for increased electric field enhancement was studied mostly experimentally, where only a few vague empirical rules were derived \cite{nanotubes_1_2_ratio_Forro}. We are aware of only one theoretical study of field enhancement for various geometries~\cite{electrostatic_screening}. Part of the reason for the lack of a solid theoretical framework is due to the absence of  solvers that can provide accurate solutions of the electric field for relatively complex geometries with a large number of boundary elements.  In techniques such as finite element or finite difference, the problem is numerically prohibitive, while for most BEM implementations, the memory requirement is often too large. The Robin Hood techniques offers a possible solution to this modeling problem.

Often the field-emission electrodes used in ion emitters are covered with extremely thin objects, such as carbon nano-tubes. Due to the presence of such very sharp tips, the electric field is extremely enhanced compared to a smooth electrode, up to factors of $10^5$.  Such enhancement can be tempered by field screening caused when other high field regions are brought to its vicinity, which is typically the case in producing such patterned electrodes. A greater density of such surface deformations stems from the need of large emission currents. Therefore, there exists a tradeoff between the field enhancement on the tip and number of tips per surface area~\cite{field_emission_exponential_current, FN_equation}.   To study the relationship between field and current enhancement, we explore two possible geometries.

\subsection{Egg-Carton Geometry}

\begin{figure}[htb]
\begin{center}
{
\includegraphics[width=0.9\linewidth,clip=true]{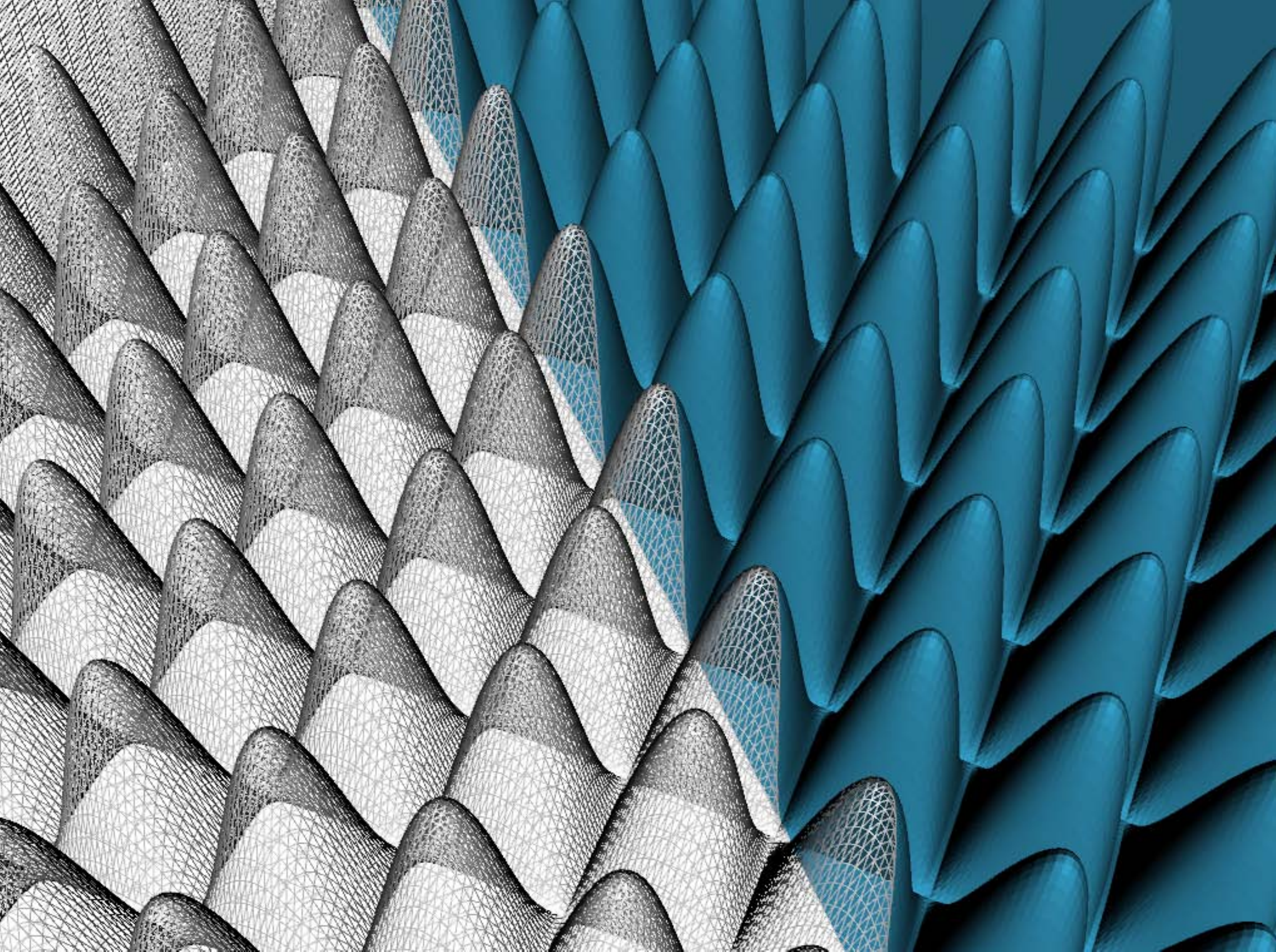}
}
\end{center}
\caption{Depiction of the egg-carton geometry. The realistic grid of boundary elements (triangles) that is used in the calculation is visible. The peak-to-spacing ratio shown for this figure is 4 to 1. }
\label{eggbox_geometry}
\end{figure}

Field enhancement on corrugates surfaces is commonly used today in surface-enhanced Raman scattering (SERS)~\cite{SERS}, where such effects can increase the measured Raman signal up to 11 orders of magnitude due to the $E^4$ dependence on the electric field strength. SERS is in principle a time-dependent phenomenon, but for certain cases, a quasi-static calculation is sufficient to estimate the quality of the corrugated surface for the enhancement of the signal. Some other aspects of field enhancement on the corrugated surface can include unusual and unexpected phenomena, such as X-ray emission~\cite{scotch_tape,bopersson}. Because of the typical nano-structured geometry that appear in some of these applications, we have chosen a simple egg-carton structure. The surface is modeled by a $|\cos{(\lambda r)}|$ function where the amplitude is constant but the spacing --controlled by $\lambda$-- is allowed to vary.  An example of such a geometry is shown in Fig. \ref{eggbox_geometry}.

\begin{figure}[htb]
\begin{center}
{
\includegraphics[width=0.4\linewidth,clip=true]{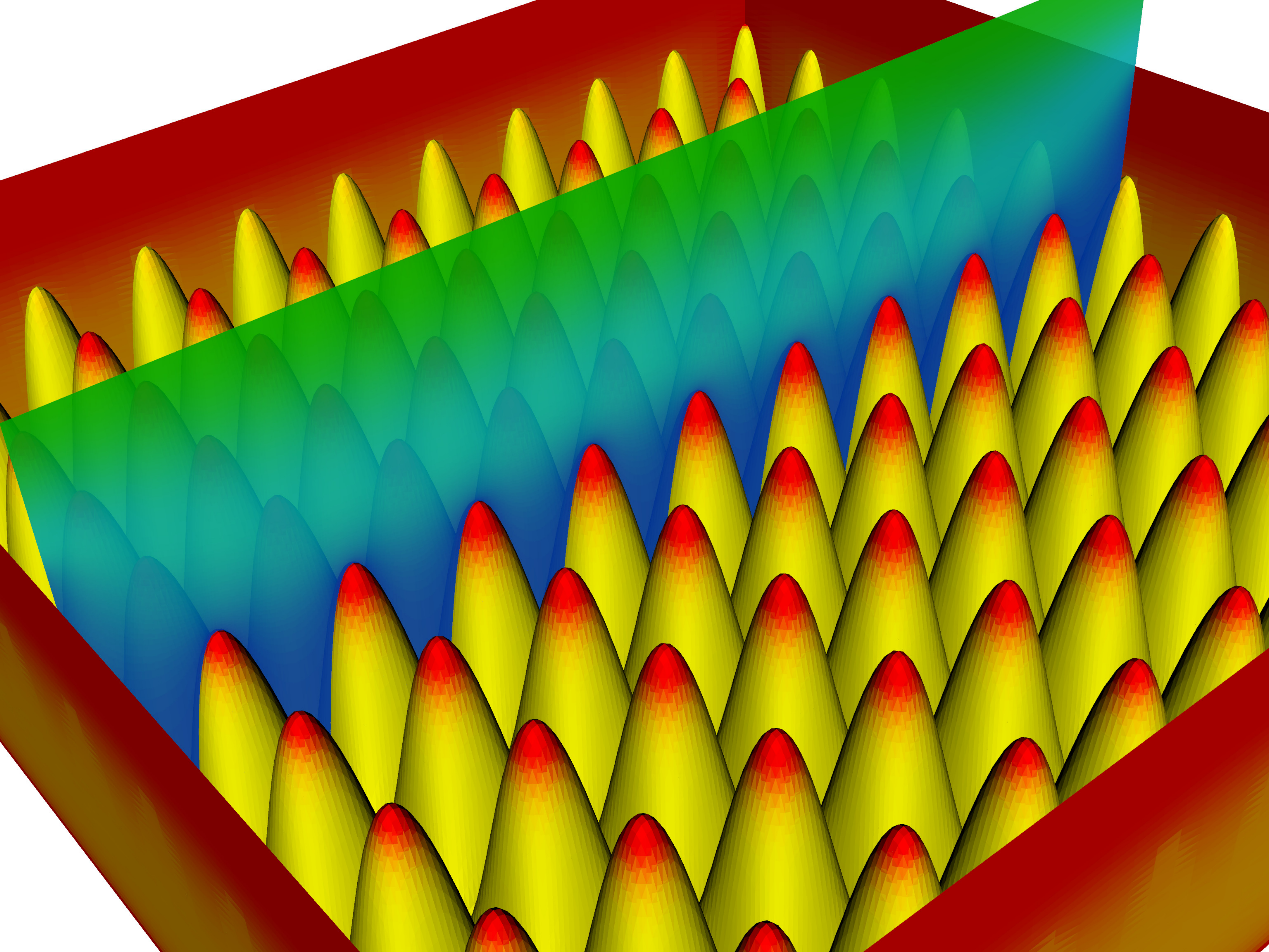}
\includegraphics[width=0.4\linewidth,clip=true]{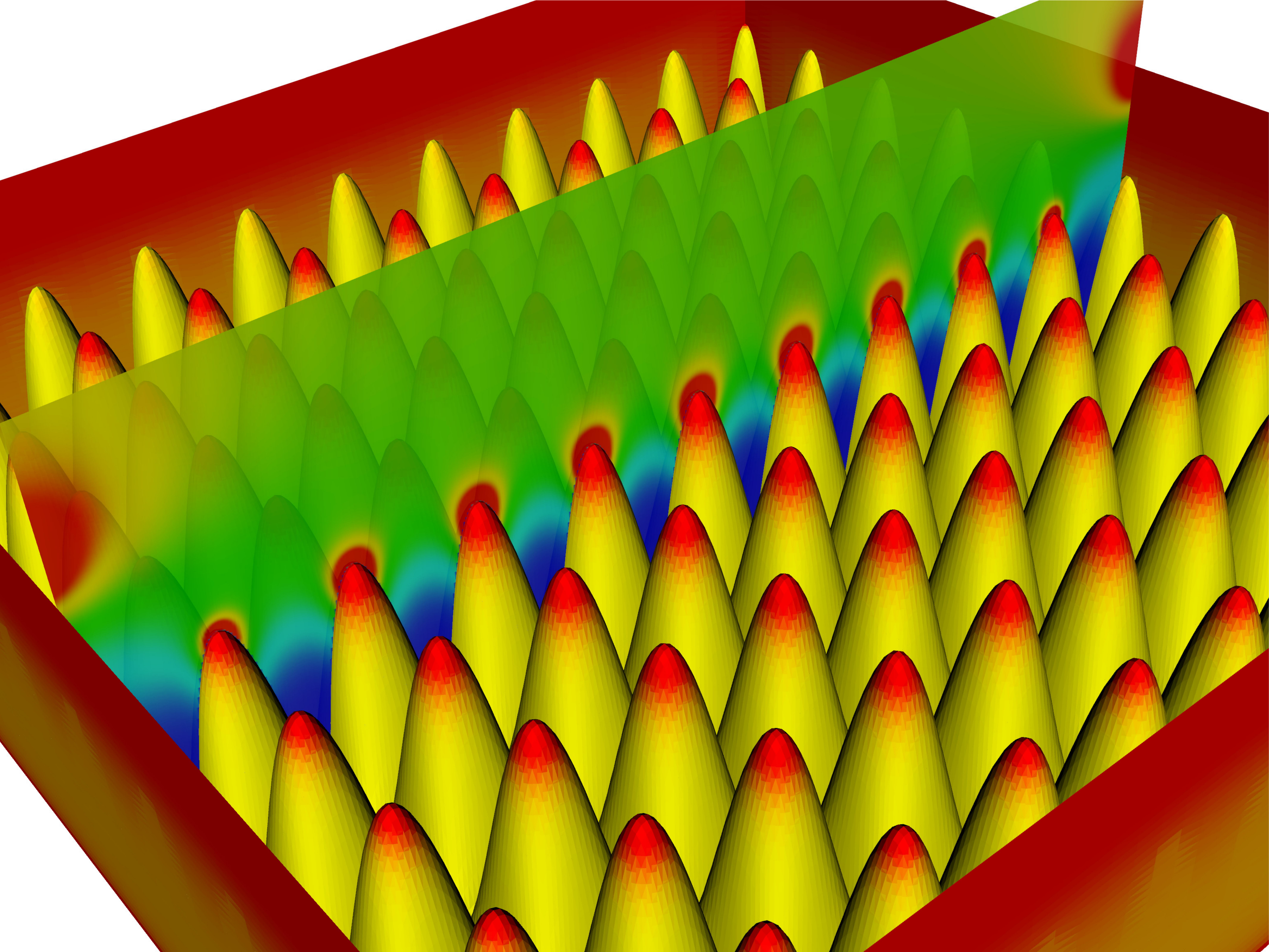}
}
\end{center}
\caption{Planes showing a cross section at which potential (left) and electric field (right) values are calculated. One can see charge accumulation around the tips.}
\label{eggbox_cuts_geo}
\end{figure}

\begin{figure}[htb]
\begin{center}
{
\includegraphics[width=0.3\linewidth,clip=true]{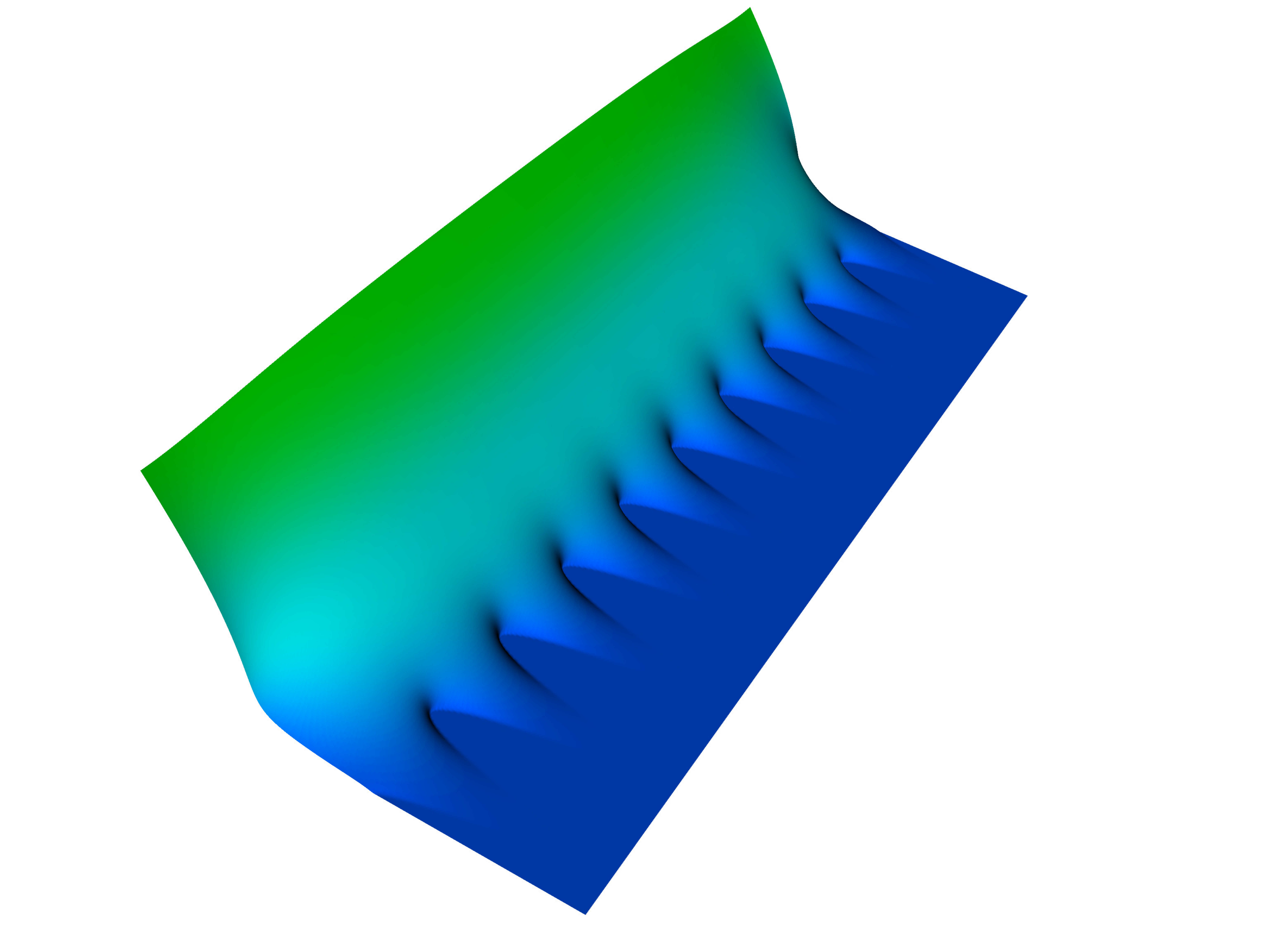}
\includegraphics[width=0.3\linewidth,clip=true]{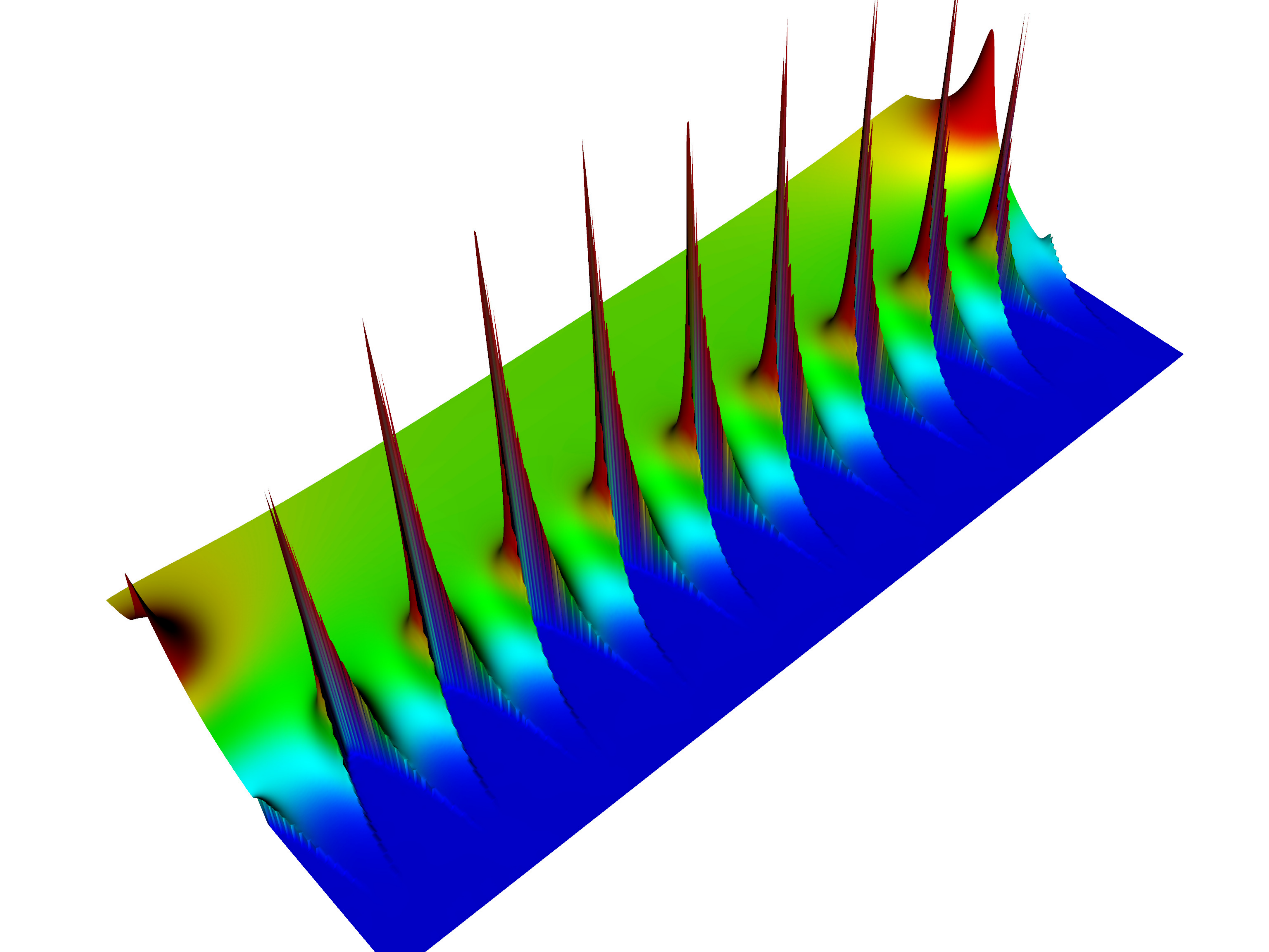}
\includegraphics[width=0.3\linewidth,clip=true]{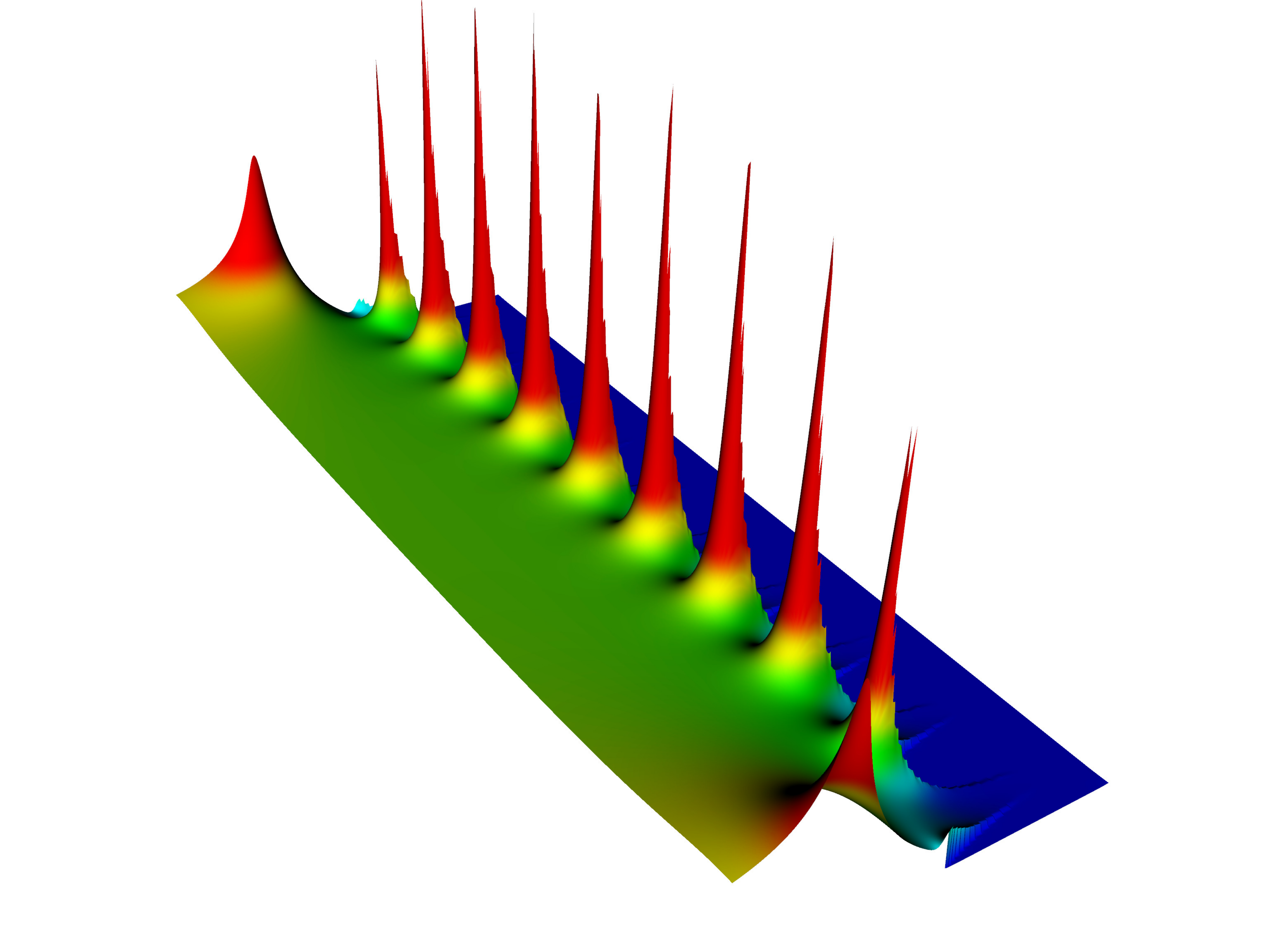}
}
\end{center}
\caption{Cross-section of the potential value - left, and the electric field strength- middle and right. Values are represented by color. The two-dimensional cross section is protruded in the direction perpendicular to the plane proportionally to the potential/field value at a given point. In this particular example, the spacing between the geometrical apexes is 4 times smaller than their amplitude. Note how quickly the edge effects disappear as one approaches the middle of the cross-section plane.}
\label{eggbox_potential_and_field}
\end{figure}

\begin{figure}[htb]
\begin{center}
{
\includegraphics[width=0.9\linewidth,clip=true]{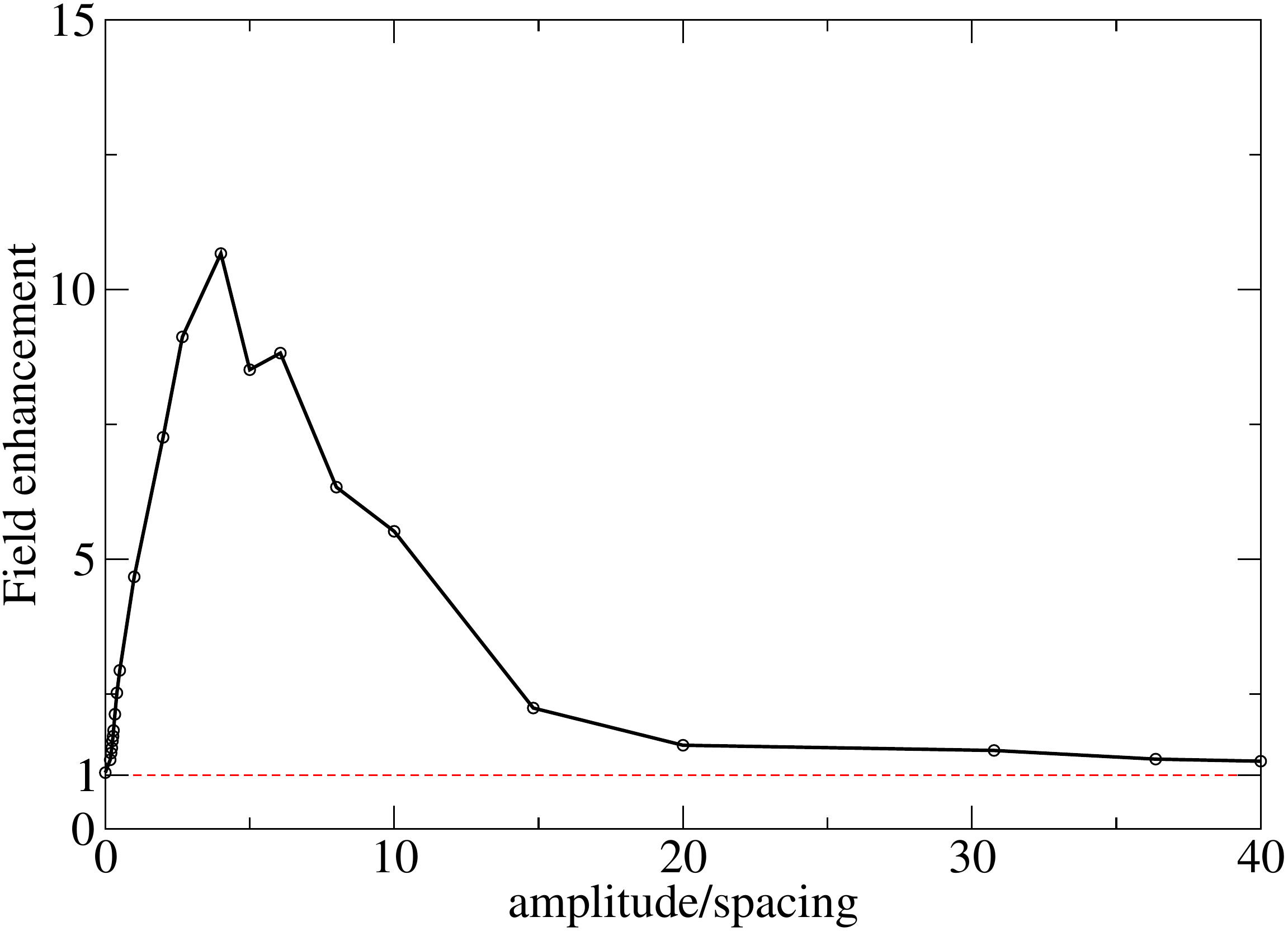}
}
\end{center}
\caption{Field enhancement values for different amplitude/spacing ratios for the egg-carton geometry.}
\label{eggbox_field_enhancement}
\end{figure}

We consider the geometry from Fig. \ref{eggbox_geometry} to be an ideal metal electrode held at a fixed voltage.  Using the Robin Hood technique outlined earlier, we find a solution that satisfies the boundary conditions at the surface of the electrode at a relative accuracy of $\bar{\epsilon} \sim 10^{-4}$. Having solved for the surface charge densities, we can then calculate cross-section of the potential and electric field, as is shown in Figures~\ref{eggbox_cuts_geo} and~Fig. \ref{eggbox_potential_and_field}.  Using the cross-section electric field values, we determine the maximum electric field strength relative to where the field is homogeneous. From this ratio, we can determine the field enhancement factor for a particular geometry. We have calculated field enhancement factor for different amplitude versus spacing ratios and results are shown in Fig.~\ref{eggbox_field_enhancement}. A ratio of 4 appears optimal, yielding a field-enhancement factor of $\sim 10$.   


What we conclude from the results in Fig. \ref{eggbox_field_enhancement} is that optimal field enhancement is achieved at a case when the amplitude is 4 times larger than spacing between spikes. Starting from a perfectly flat surface, which has field enhancement of 1, the introduction of spikes (corrugation) starts yielding field enhancement. But as soon as the density is too high, field screening onsets and begins to counteract the field enhancement.  At amplitude spacing ratio of 10, the system is almost equivalent to a smooth, flat electrode. These results show the strong, non-trivial dependence on the geometry of the system and how optimization needs to be carefully modeled. Since Maxwell's equations are scale-invariant, the geometry of the egg-carton used in this simulation applies equally at the nano scale, as well as the macro-scale without changing our conclusions. However, even minute changes in the details of the geometry (adding more spikes, varying height, etc.) will yield different behavior on optimal parameters for field enhancement. Therefore, the best option in optimizing such systems is an accurate model of the geometry.  With a parallel version of the Robin Hood method, modeling of such geometries is now possible.
 
\subsection{Ion emitters}

\begin{figure}[htb]
\begin{center}
{
\includegraphics[width=0.4\linewidth,clip=true]{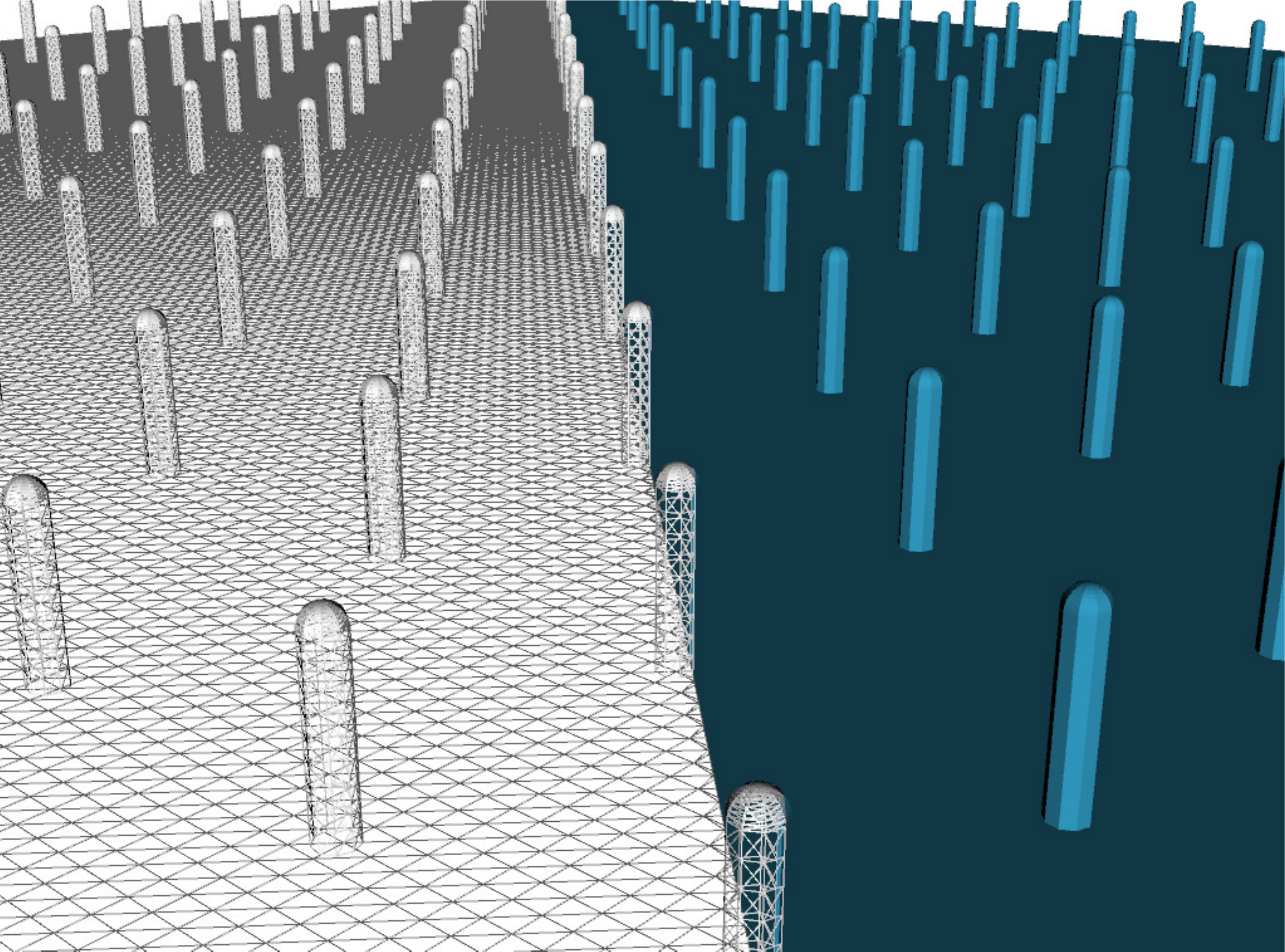}
\includegraphics[width=0.4\linewidth,clip=true]{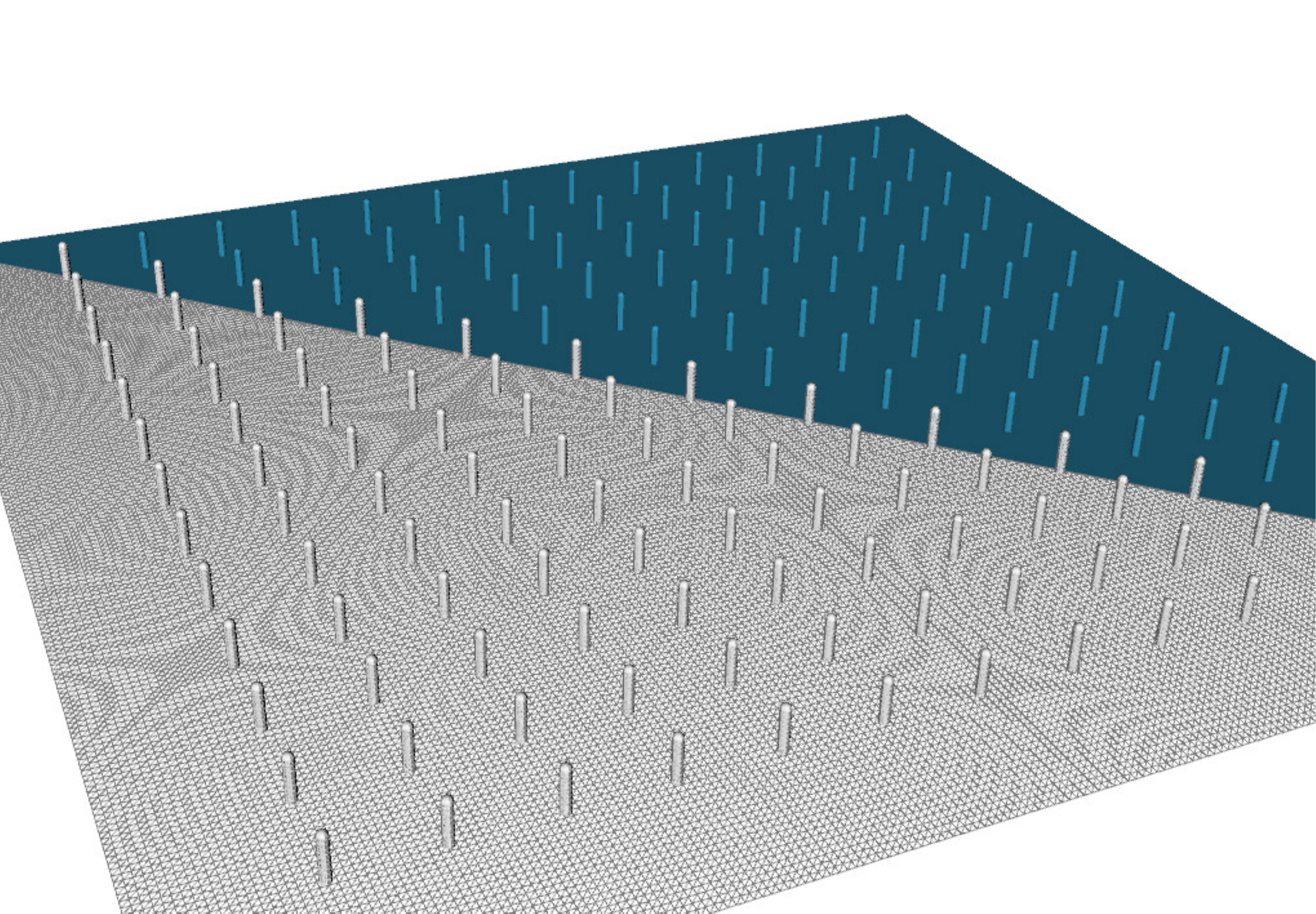}
}
\end{center}
\caption{Geometry of the nano-tube array. On one half the actual mesh of triangles used in calculation is shown.}
\label{fusion_array_geometry}
\end{figure}

Our second example is taken from the large field of research of ion emitters and gas detectors. In these systems, a neutral atom or a molecule is ionized once it approaches the electrode and is then accelerated in the electric field. Ion emitters have a wide variety of uses, such as compact neutron generators for oil well logging \cite{berkeley_guys, surface_state_enhancement}. Even though average ionization energies are relatively small -- few tens of eVs--, the ionization process is strongly determined by the shape of the potential barrier, i.e. by the electric field strength. Typical field strengths are  on the order of $10^9-10^{10}$ V/m, and achieving them with smooth, compact electrodes is difficult. Such difficulties can be overcome by having a geometry that can significantly enhance the electric field. Recently such devices have been producing using different nano structures, in particular using carbon nanotubes. Again for this study we have chosen a geometry that resembles experimental efforts, such as shown in Fig. \ref{fusion_array_geometry}.

\begin{figure}[htb]
\begin{center}
{
\includegraphics[width=0.9\linewidth,clip=true]{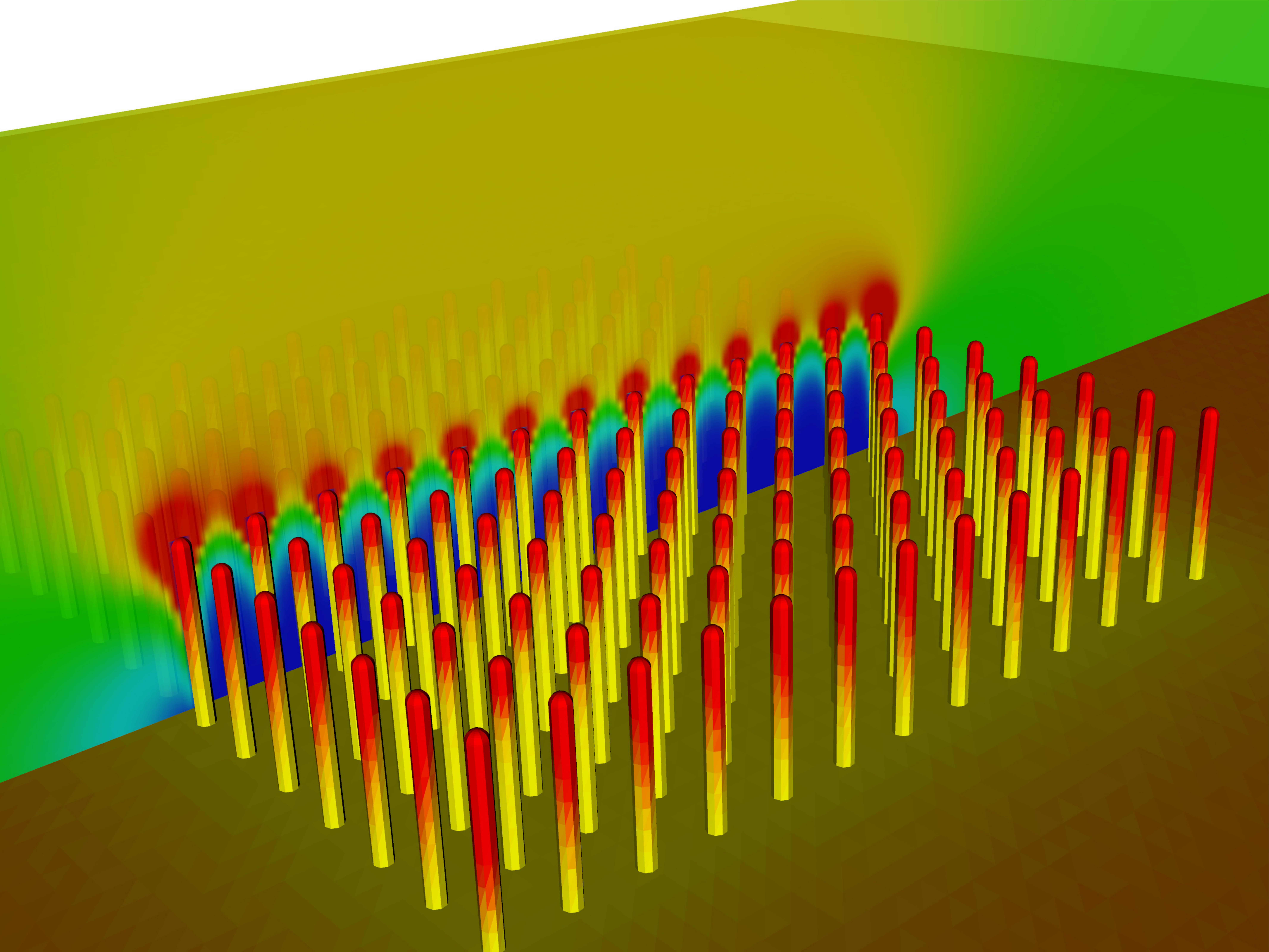}
}
\end{center}
\caption{Cross section plane showing the electric field values. On the nano-tubes the charge density values are shown by means of color. Red color means large charge density, and one can see that the last row of nano-tubes is slightly more charged than the internal ones, having as the maximally charged nano-tube the one on the corner - this is also reflected in the electric field values in the cross-section plane.}
\label{fusion_array_slice_geo}
\end{figure}

Following closely the previous example, we calculate different arrangements of the nanotube array by varying both the radius of the nanotube and the inter-spacing between tubes. Once the surface charge densities are obtained, we calculate the potential and field values in a cross section plane, as shown in Fig.~\ref{fusion_array_slice_geo}.

%

\begin{figure}[htb]
\begin{center}
{
\includegraphics[width=0.4\linewidth,clip=true]{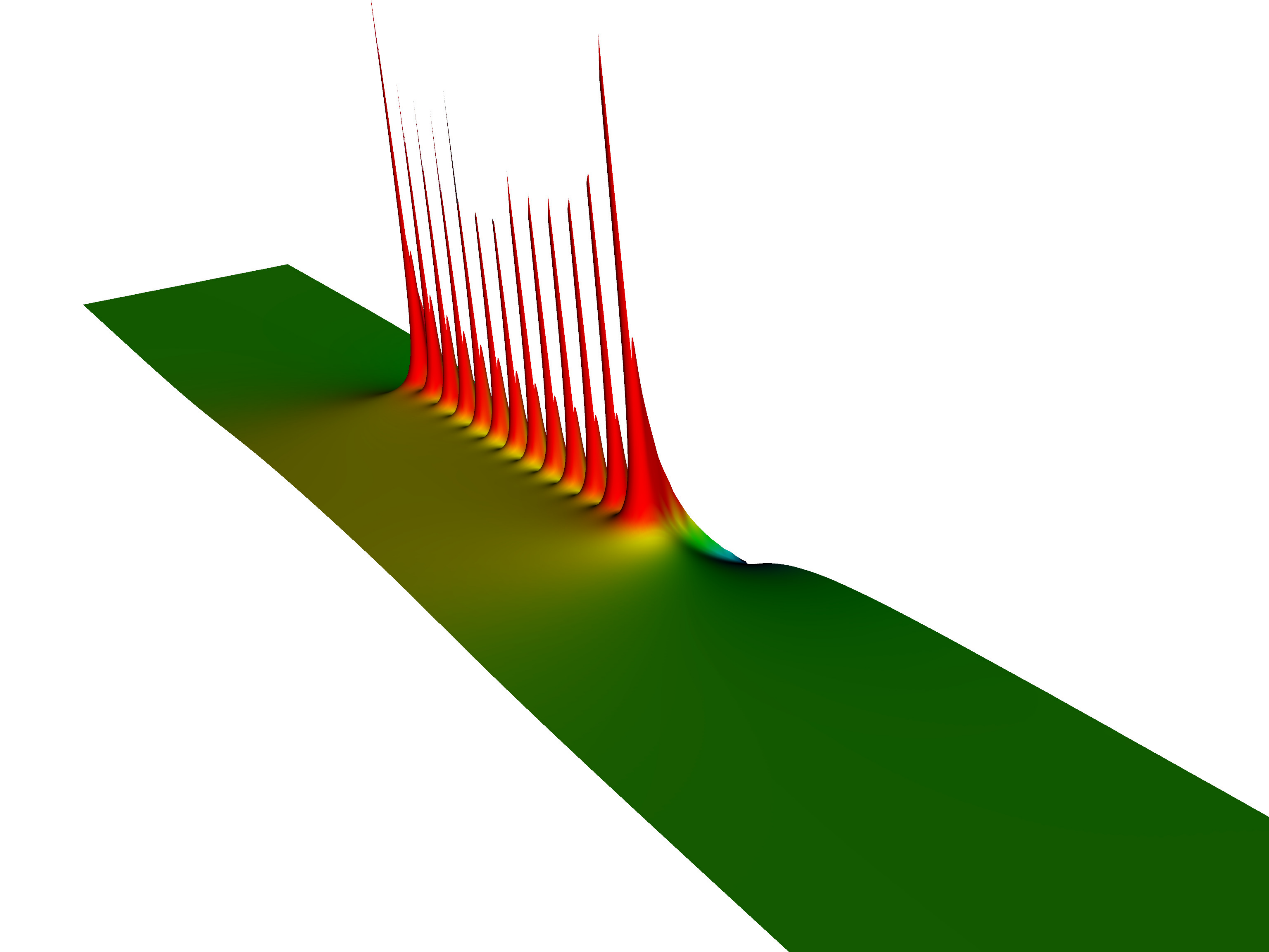}
\includegraphics[width=0.4\linewidth,clip=true]{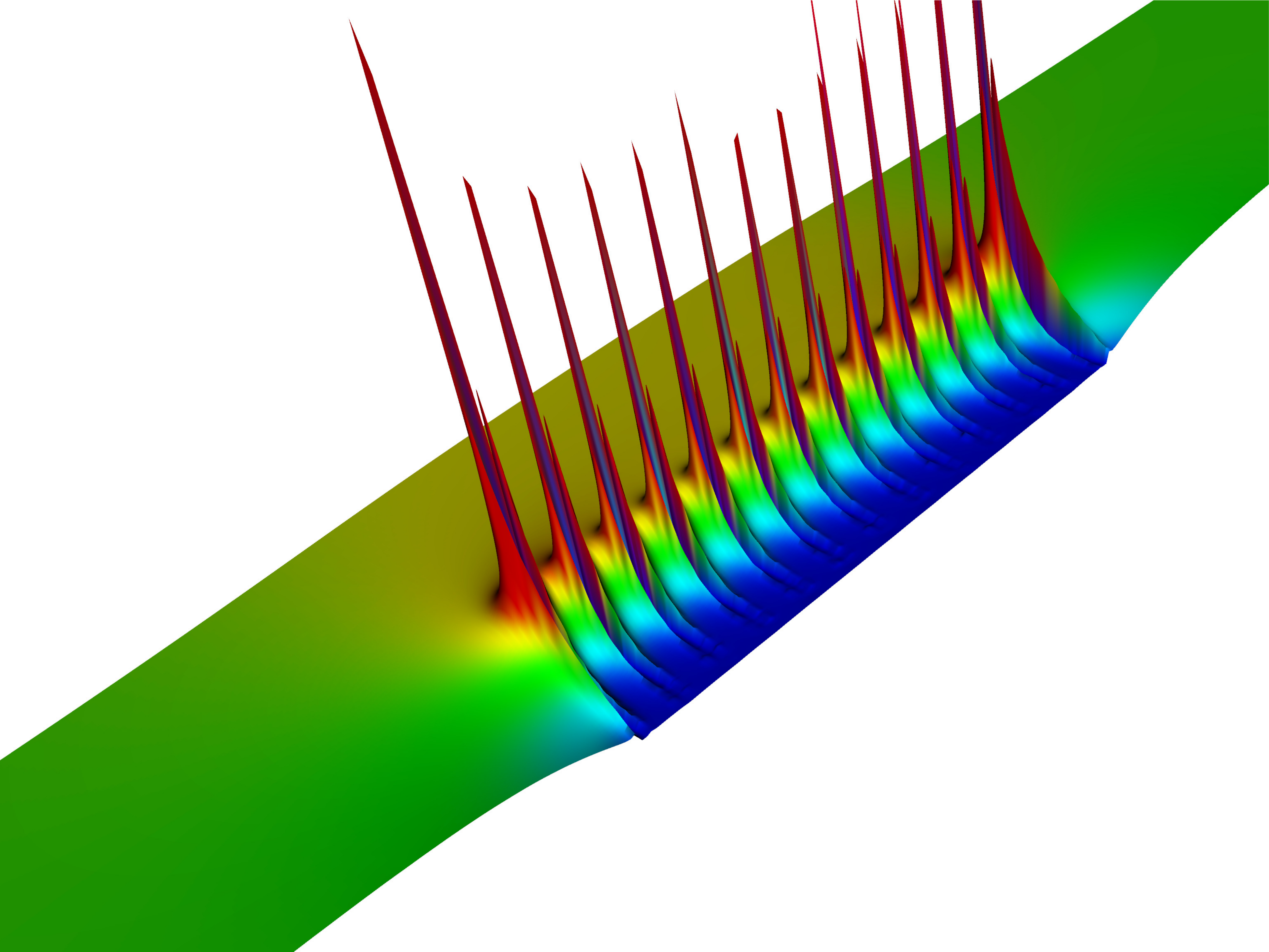}
}
\end{center}
\caption{Electric field value in a cross-section plane shown from two different angles for the nanotube geometry. The cross-section plane is flat and it is being extruded in the perpendicular direction proportionally to the field value.The larger field enhancement at the very end of the array is due to finite size effects and are not taken into account in the the field enhancement calculation.}
\label{fusion_array_slice_field}
\end{figure}

\begin{figure}[htb]
\begin{center}
{
\includegraphics[width=0.9\linewidth,clip=true]{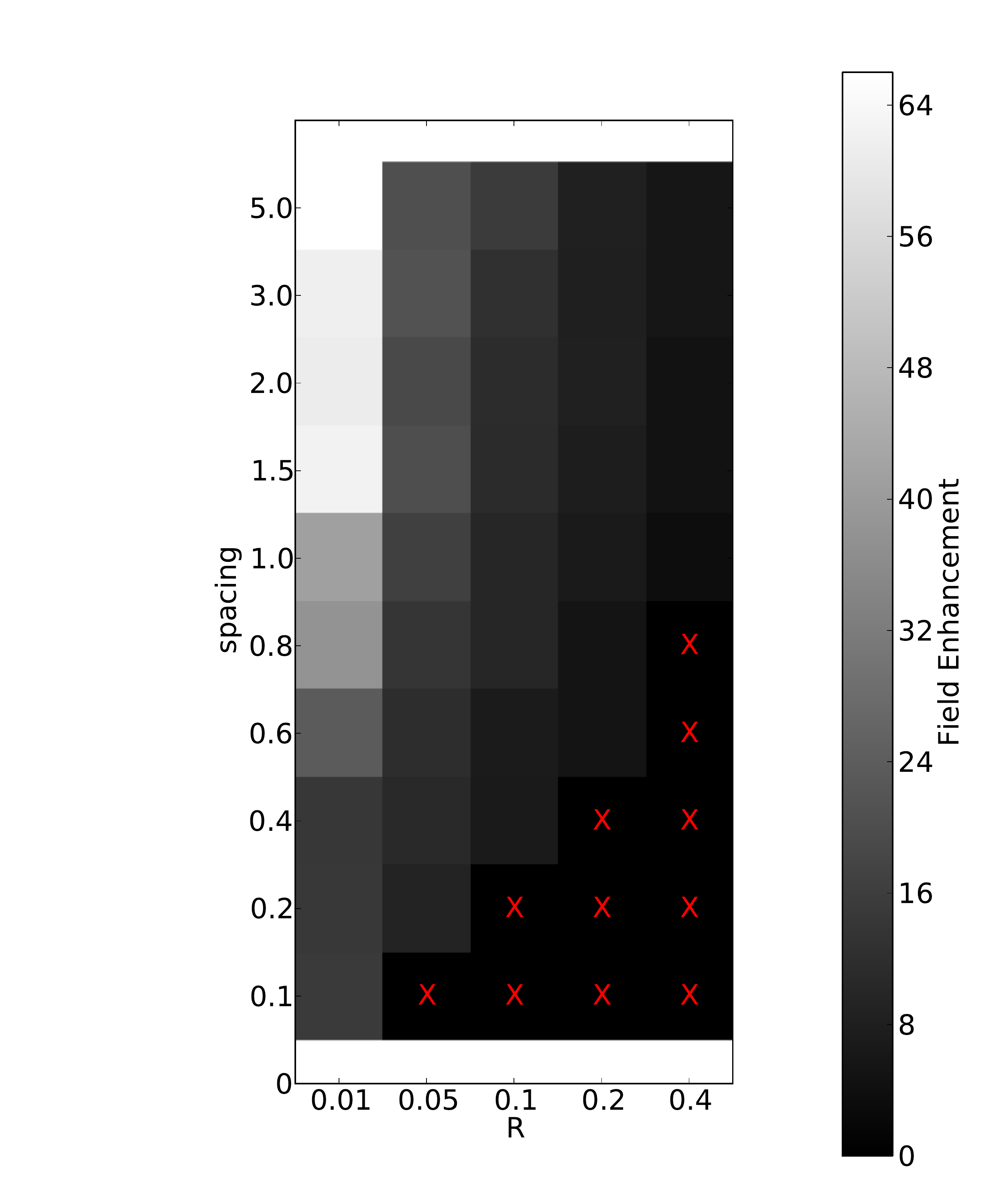}
}
\end{center}
\caption{Field enhancement values depending on the radius and inter-spacing of the nano-tube array. The red X's show impossible configurations where nano-tubes would intersect one other. Notice that simple scaling is not possible in this case since the height of the nano-tube is kept constant in all cases.  Units are arbitrary.}
\label{fusion_array_field_enhancement}
\end{figure}

Having all the data for the field enhancement as was done for the egg-carton example, we obtain the results shown in Fig. \ref{fusion_array_field_enhancement}. The maximum field enhancement factor found using this nanotube structure can be as high as $\times 64$ (for a nanotube with radius-to-inter-spacing value of 500).

The results for field enhancement in Fig. \ref{fusion_array_field_enhancement} once again demonstrate that one isolated nanotube yields largest field enhancement. A higher packing density creates a screening effect sets which reduces this enhancement. In order to optimize the current per surface area, one should couple these results with a model of current strength dependence on the electric field strength.  This approach is beyond the scope of this paper, but the interested reader should consult~\cite{FN_equation}. Again, in more realistic cases, the nanotubes may possess variations in orientation and size that may alter the results shown here. None of these features poses a problem in principle, as long as one can provide a 3D mesh that describes the desired geometry with sufficient accuracy.

\section{Summary and Future Work}
\label{sec:conclusion}

In conclusion, we have demonstrated the versatility of the Robin Hood technique across a variety of scales.  In the case of large scale structures, such as the KATRIN detector, accuracy can be retained by introducing a high degree of segmentation and taking advantage of the low memory footprint used by the technique.  In the case of micro-scale structures, we have shown that Robin Hood is capable of calculating large periodic arrays of complex structures. Microstructuring of electrodes is playing increasingly important role in many current applications~\cite{graetzel}. Simulations such as Robin Hood allow to move beyond simple approximations and provide more accurate optimizations.

In principle, the boundary element method in conjunction with the Robin Hood technique can be made to solve {\em any} positive-definite matrix that exibits linear dependence.  As such, we believe the technique can be utilized in other contexts, such as time-varying electromagnetic fields and linearized gravity.  Such extensions are the subject of further inquiry.

\ack
J. A. Formaggio is supported by the United States Department of Energy under Grant No. DE-FG02-06ER-41420.  The authors wish to thank the KATRIN collaboration for their support.


\section*{Appendix A: Potential from Triangular Surfaces}

\begin{figure}[htb]
\begin{center}
{
\includegraphics[width=0.45\linewidth,clip=true]{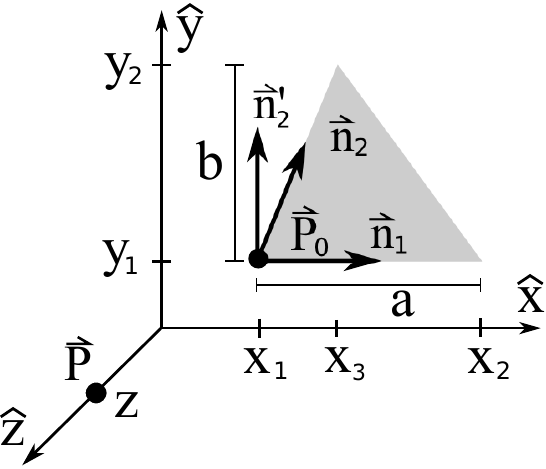}
}
\end{center}
\caption{A triangular surface defined by a corner point $\vec{P}_{0}$, the length of the longest side $a$ and corresponding height $b$, and the unit vectors in the directions of the sides connected to $\vec{P}_{0}$, labeled $\vec{n}_{1}$ and $\vec{n}_{2}$ ($\vec{n}_{1}$ always points in the direction of $a$).  The field point is defined as $\vec{P}$, with local coordinates $(0,0,z)$.  The corners of the triangle are recast into local coordinates to facilitate integration.}
\label{potentialFromTriangle}
\end{figure}

Let a triangular surface be defined by a corner $\vec{P}_{0}$, the lengths of its sides $a$ and $b$, and the unit vectors defining its sides $\vec{n}_{1}$ and $\vec{n}_{2}$ (see Fig. \ref{potentialFromTriangle}).  During the computation of the electric potential and field at a field point $\vec{P}$, it is convenient to transform into a local coordinate frame\footnote{as recommended in \cite{Birtles}.}, where the triangle lies in the $\hat{x}$-$\hat{y}$ plane and the field point lies along the $\hat{z}$-axis, as depicted in Figure \ref{potentialFromTriangle}.  In the local coordinate system, the parameters necessary for the potential calculation are:

\begin{eqnarray}
z & = & \left( \vec{P}- \vec{P}_{0} \right) \cdot \vec{n}_{3}, \\
x_{1} & = & \left( \vec{P}_{0} - \vec{P} \right) \cdot \vec{n}_{1}, \\
x_{2} & = & x_{1} + a, \\
x_{3} & = & x_{1} + (\vec{n}_{1} \cdot \vec{n}_{2}) \cdot b, \\
y_{1} & = & \frac{z}{|z|} \cdot \left( \vec{P}_{0} - \vec{P} \right) \cdot \vec{n}^{\prime}_{2}, \\
y_{2} & = & y_{1} + \frac{z}{|z|} \cdot b,
\end{eqnarray}
where $\vec{n}_{3} = \vec{n}_{1} \times \vec{n}_{2}$, and $\vec{n}^{\prime}_{2}$ is the unit normal to $\vec{n}_{1}$ in the direction of $\vec{n}_{2}$ ($\vec{n}^{\prime}_2 = \frac{\vec{n}_{2}-(\vec{n}_{2} \cdot \vec{n}_{1})\vec{n}_1}{|\vec{n}_{2}-(\vec{n}_{2} \cdot \vec{n}_{1})\vec{n}_{1}|}$).

The analytic calculation of the potential from this triangular surface (assuming a constant charge density $\sigma$) can be found by directly evaluating the following integral: 
\begin{equation}
V = \frac{\sigma}{4 \pi \epsilon_{0}} \int_{y_{1}}^{y_{2}} \int_{a_1 + b_1 y}^{a_2 + b_2 y} \frac{1}{\sqrt{x^{2}+y^{2}+z^{2}}} \cdot dx \cdot dy,
\label{potentialFromTriangle1}
\end{equation}
where $a_{1} = \frac{x_{1}y_{2} - x_{3}y_{1}}{(y_{2}-y_{1})}$, $a_{2} = \frac{x_{2}y_{2} - x_{3}y_{1}}{(y_{2}-y_{1})}$, $b_{1} = \frac{x_{3}-x_{1}}{y_{2}-y{1}}$ and $b_{2} = \frac{x_{3}-x_{2}}{y_{2}-y{1}}$.  

We first manipulate the equation into a more soluble form: 
\begin{equation}
V = \frac{\sigma}{4 \pi \epsilon_0} \int_{y_{1}}^{y_{2}} dy \int_{u_{1}}^{u_{2}} \frac{\cosh{(u)} \cdot du}{\sqrt{\sinh^2{(u)}+1}}, 
\label{firstIntegral2}
\end{equation}
where $u_{1} = \sinh^{-1}{\left(\frac{a_1 + b_1 y}{\sqrt{y^2+z^2}}\right)}$ and $u_{2} = \sinh^{-1}{\left(\frac{a_2 + b_2 y}{\sqrt{y^2+z^2}}\right)}$.  Using the identity
\begin{equation}
\cosh^2(x)-\sinh^2(x)=1,
\label{coshsinhIdentity}
\end{equation}
Equation \ref{firstIntegral2} becomes
\begin{eqnarray}
V &= \frac{\sigma}{4 \pi \epsilon_0} \int_{y_{1}}^{y_{2}} dy & \int_{u_{1}}^{u_{2}}du = \nonumber \\
&= \frac{\sigma}{4 \pi \epsilon_0} \int_{y_{1}}^{y_{2}} dy & \left[\sinh^{-1}\left( \frac{a_2+b_2 y}{\sqrt{y^2+z^2}}\right) - \right. \nonumber \\
&& \left.  \sinh^{-1}\left( \frac{a_1 + b_1 y}{\sqrt{y^2+z^2}}\right) \right].
\label{firstIntegral3}
\end{eqnarray}

After dividing by $z$ to make the integral dimensionless, Equation \ref{firstIntegral3} becomes
\begin{eqnarray}
V &= \frac{\sigma}{4 \pi \epsilon_{0}} \cdot z \cdot  &\left[ \int_{u_{1}}^{u_{2}}du\cdot \sinh^{-1} \left( \frac{a_2^{\prime} + b_2 u}{\sqrt{u^{2}+1}}\right) - \right. \nonumber \\
&&\left. \int_{u_{1}}^{u_{2}}du\cdot \sinh^{-1} \left( \frac{a_{1}^{\prime} + b_1 u}{\sqrt{u^{2}+1}} \right) \right],
\label{potentialFromTriangle2}
\end{eqnarray}
where $a_{1}^{\prime} = \frac{a_{1}}{z}$, $a_{2}^{\prime} = \frac{a_{2}}{z}$, $u = \frac{y}{z}$ and $u_{i} = \frac{y_{i}}{z}$.  

If we define the following indefinite integral
\begin{equation}
  I_{1}(a,b,u) = \int \sinh^{-1} \left(\frac{a + b u}{\sqrt{u^{2}+1}}\right) \cdot du,
\label{potentialFromTriangle3a}
\end{equation}
Equation \ref{potentialFromTriangle2} can be rewritten as
\begin{eqnarray}
V &= \frac{\sigma}{4 \pi \epsilon_0} \cdot z \cdot & \left[ I_1(a_2^\prime,b_2,u_2) - I_1(a_2^\prime,b_2,u_1)- \right. \nonumber \\
&& \left.I_1(a_1^\prime,b_1,u_2) + I_1(a_1^\prime,b_1,u_1)\right].
\label{potentialFromTriangle4}
\end{eqnarray}

\subsection{Solution for $I_1(a,b,u)$}

The solution to $I_1(a,b,u)$ presented here is based on the technique outlined in \cite{Birtles}.  $I_1$ is first integrated by parts: 
\begin{equation}
I_{1}(a,b,u) = F_{1}(a,b,u) + I_{3}(a,b,u) - I_{4}(a,b,u),
\end{equation}
where
\begin{eqnarray}
\label{F1}
F_{1}(a,b,u) &=& u \cdot \sinh^{-1}\left(\frac{a + b u}{\sqrt{u^2+1}}\right), \\
I_{3}(a,b,u) &=& \int \frac{a u^2\cdot du}{U^{\frac{1}{2}}(u^2+1)}, \\
I_{4}(a,b,u) &=& \int \frac{b u\cdot du}{U^{\frac{1}{2}}(u^2+1)},
\end{eqnarray}
and the substitution
\begin{equation}
U = \left((1 + a^2) + 2 (a b) u + (1+b^2)u^2\right)
\end{equation}
has been made merely to clarify the equations.   By noting that $I_{3}$ can be rewritten as
\begin{equation}
I_{3}(a,b,u) = a\cdot \left( \int \frac{du}{U^{\frac{1}{2}}} - \int \frac{du}{U^{\frac{1}{2}}(u^2+1)}\right),
\end{equation}
we can move the second term into $I_{4}$, leaving us with
\begin{equation}
I_{1}(a,b,u) = F_{1}(a,b,u) + \tilde{I}_{3}(a,b,u) - \tilde{I}_{4}(a,b,u),
\end{equation}
where
\begin{eqnarray}
\tilde{I}_{3}(a,b,u) &=& \int \frac{a \cdot du}{U^{\frac{1}{2}}}, \\
\tilde{I}_{4}(a,b,u) &=& \int \frac{\left(b u+a\right)\cdot du}{U^{\frac{1}{2}}(u^2+1)}.
\end{eqnarray}


By affecting a change in variables,
\begin{eqnarray}
a^{\prime} &=& 1+b^2, \nonumber \\
b^{\prime} &=& 2 a b, \nonumber \\
c^{\prime} &=& 1+a^2,
\end{eqnarray}
 $\tilde{I}_{3}$ can be cast into a form readily found in tables of integrals: 
\begin{equation}
\frac{\tilde{I}_{3}}{a} = \int \frac{ du}{\sqrt{a^{\prime} u^2 + b^{\prime} u + c^{\prime}}}.
\end{equation}
The general solution to this integral for $a^{\prime}>0$ is taken from \cite{Hudson} to be
\begin{eqnarray}
\int \frac{ du}{\sqrt{a^{\prime} u^2 + b^{\prime} u + c^{\prime}}} =\qquad \qquad \qquad \qquad \qquad \qquad &&\nonumber \\
 \frac{1}{\sqrt{a^{\prime}}} \ln\left(2 a^{\prime} u + b^{\prime} + 2 \sqrt{a^{\prime}} \sqrt{a^{\prime} u^2 + b^{\prime} u + c^{\prime}} \right). &&
\label{oneoverradical}
\end{eqnarray}
Recast into our variables of interest, we arrive at an equation for $\tilde{I}_{3}$:
\begin{eqnarray}
\tilde{I}_{3}(a,b,u) &= \frac{a}{\sqrt{b^2+1}} \cdot \ln \left(2 \left(b (a+b u)+u + \right. \right. \qquad \qquad \qquad \\
&\left. \left. \sqrt{b^2+1} \sqrt{a^2+2 a b u+\left(b^2+1\right) u^2+1}\right)\right). \nonumber 
\end{eqnarray}


Integral $\tilde{I}_{4}$ can be converted into a form with an analytic solution by performing the following change of variables:
\begin{eqnarray}
a^{\prime} &=& 1+b^2, \nonumber \\
b^{\prime} &=& a b, \nonumber \\
c^{\prime} &=& 1+a^2,\nonumber \\
h&=& b, \nonumber \\
k&=& a.
\end{eqnarray}
The recast integral is then
\begin{equation}
\tilde{I}_{4} = \int \frac{(h   u + k) d   u}{(  u^2+1)\sqrt{a^{\prime}   u^2 + 2b^{\prime}   u + c^{\prime}}},
\label{bowmanRecast}
\end{equation}
and can be solved using a technique outlined in \cite{Bowman} by making the substitutions
\begin{equation}
  u \rightarrow \frac{\lambda t + 1}{t - \lambda}, \: d   u \rightarrow -\frac{(\lambda^2+1)}{(t-\lambda)^2}dt
\end{equation}
in Eq. \ref{bowmanRecast} and recovering the form
\begin{equation}
\tilde{I}_4 = \int \frac{(l t + m) d t}{(t^2+1)\sqrt{\alpha t^2 + 2\beta t + \gamma}}.  
\label{secondform}
\end{equation}
This can be accomplished by making the following substitutions: 
\begin{eqnarray}
l =&-h\lambda - k & = -b \lambda - a, \nonumber \\
m =& k \lambda - h &= a \lambda -b, \nonumber \\
\alpha =& \lambda^2 a^{\prime} + 2 b^{\prime} \lambda + c^{\prime} =& \nonumber \\
=& (1+b^2)\lambda^2 + 2 (a b) \lambda + (1+a^2), \nonumber \\
\beta =& -b^{\prime} \lambda^2 + (a^{\prime}-c^{\prime})\lambda + b^{\prime}, \nonumber \\
\gamma =& a^{\prime} - 2 b^{\prime} \lambda + c^{\prime} \lambda^2 =&\nonumber \\
=& (1+b^2) - 2 (a b) \lambda + (1+a^2) \lambda^2.
\end{eqnarray}
Now, if we let $\lambda$ equal one of the roots of $\beta$,
\begin{eqnarray}
\lambda &=& \frac{(a^{\prime}-c^{\prime})\pm\sqrt{(c^{\prime}-a^{\prime})^2+4b^{\prime 2}}}{2 b^{\prime}} =\\
&=& \left(\frac{b}{a}\right) {\rm~or~} \left(-\frac{a}{b}\right),
\end{eqnarray}
we can remove $\beta$ from Equation \ref{secondform} and split the integral into two parts:
\begin{equation}
\tilde{I}_4 = \int \frac{l t  d t}{(t^2+1)\sqrt{\alpha t^2 + \gamma}} + \int \frac{m d t}{(t^2+1)\sqrt{\alpha t^2 + \gamma}} .
\label{twointegrals}
\end{equation}
The second integral can be put in the same form as the first if we affect another change of variables, $t \rightarrow \frac{1}{s}$, and take $t>0$, $s>0$:
\begin{equation}
\tilde{I}_4 = \int \frac{l t  d t}{(t^2+1)\sqrt{\alpha t^2 + \gamma}} - \int \frac{m s d s}{(s^2+1)\sqrt{\gamma s^2 +\alpha}} .
\label{twointegrals_invert}
\end{equation}
Solutions to integrals of this form are defined as
\begin{equation}
\int \frac{t d t}{(t^2+1)\sqrt{\xi t^2 + \zeta}} = \frac{\tan ^{-1}\left(\sqrt{\frac{\xi t^2+\zeta}{\xi-\zeta}}\right)}{\sqrt{\xi-\zeta}}.
\label{atanintegral}
\end{equation}
Using Equation \ref{atanintegral} and taking the more negative root of $\beta$ ($\lambda = -\frac{a}{b}$), we arrive at a closed form solution for $\tilde{I}_{4}$: 
\begin{equation}
\tilde{I}_{4}(a,b,u) = -(\frac{a^2}{b} + b) \cdot \frac{1}{\sqrt{\gamma-\alpha}} \cdot \tan^{-1}\left( \sqrt{\frac{\gamma t^2 + \alpha}{\gamma - \alpha}}\right),
\label{I4positivelambda}
\end{equation}
where
\begin{eqnarray}
\alpha &=& 1+ \frac{a^2}{b^2},\nonumber \\
\gamma &=&\frac{\left(a^2+b^2\right) \left(a^2+b^2+1\right)}{b^2},\nonumber \\
t &=&-\left(\frac{b u+a}{a u-b}\right).
\label{I4positivelambdaeqns}
\end{eqnarray}
For computation, it is often beneficial to use an analytic form for $\tilde{I}_4(a,b,u_2) - \tilde{I}_4(a,b,u_1)$ to avoid intermediate computations that may be complex: 
\begin{eqnarray}
\tilde{I}_{4}(a,b,u_{2}) - \tilde{I}_{4}(a,b,u_{1}) =  \qquad \qquad \qquad \qquad \qquad \qquad &&  \\
=\left( \left(\frac{a^2}{b}+b\right) \cdot \frac{1}{\sqrt{\gamma-\alpha}} \right) \times \qquad \qquad \qquad \qquad \qquad  && \nonumber \\
\tan^{-1}\left(\frac{\sqrt{\gamma-\alpha}\cdot \left(\sqrt{\gamma t_{2}^{2}+\alpha} -\sqrt{\gamma t_{1}^{2}+\alpha}\right)} {(\gamma-\alpha) + \sqrt{\gamma t_2^2+\alpha\cdot \sqrt{\gamma t_1^2 + \alpha}}}\right) .&& \nonumber
\label{I4diff2}
\end{eqnarray}
It is also worth noting that, if the integration crosses a divergence in the integrand of $\tilde{I}_4$ (if $u_1<\frac{b}{a} < u_2$ or $u_1>\frac{b}{a} > u_2$), it is necessary to perform a branch cut on the interval of integration.  


When $z=0$, the above formalism cannot be applied, since it would involve a division by zero.  Instead, we start with a modified version of Equation \ref{firstIntegral3}:
\begin{eqnarray}
V &= \frac{\sigma}{4 \pi \epsilon_0} \int_{y_{1}}^{y_{2}} dy & \int_{u_{1}}^{u_{2}}du = \nonumber \\
&= \frac{\sigma}{4 \pi \epsilon_0} \int_{y_{1}}^{y_{2}} dy & \left[\sinh^{-1}\left( \frac{a_2+b_2 y}{|y|}\right) - \right. \nonumber \\
&& \left.  \sinh^{-1}\left( \frac{a_1 + b_1 y}{|y|}\right) \right].
\label{zEqualsZero1}
\end{eqnarray}
These two integrals can be made more soluble by integrating by parts:
\begin{eqnarray}
V &= \frac{\sigma}{4 \pi \epsilon_0} &\left[ y\left(\sinh^{-1}\left( \frac{a_2+b_2 y}{|y|} \right) - \right. \right. \nonumber \\
&&\left. \left. \sinh^{-1}\left( \frac{a_1+b_1 y}{|y|} \right)\right|^{y_2}_{y_1} +\right. \nonumber \\
&& \left. \int_{y_{1}}^{y_{2}} \frac{dy}{\sqrt{(1+b_2^2)y^2+(2a_2 b_2 )y+a_2^2}} - \right. \nonumber \\
&& \left. \int_{y_{1}}^{y_{2}} \frac{dy}{\sqrt{(1+b_1^2)y^2+(2a_1 b_1 )y+a_1^2}} \right].
\label{zEqualsZero2}
\end{eqnarray}
The integrals in Equation \ref{zEqualsZero2} can be performed using the identity in Equation \ref{oneoverradical}.

\end{document}